\documentclass[journal,twoside]{asmb}

\begin{document}
\title{Opportunistic Relay in Multicast Channels with Generalized Shadowed Fading Effects: A Physical Layer Security Perspective}

\author[1]{S. M. S. Shahriyer}
\author[2]{A. S. M. Badrudduza}
\author[3]{S. Shabab}
\author[4]{M. K. Kundu}
\author[5]{Heejung Yu}

\affil[1,2]{Department of Electronics \& Telecommunication Engineering, Rajshahi University of Engineering \& Technology (RUET), Rajshahi-6204, Bangladesh}
\affil[3]{Department of Electrical \& Electronic Engineering, RUET}
\affil[4]{Department of Electrical \& Computer Engineering, RUET}
\affil[5]{Department of Electronics and Information Engineering, Korea University, Sejong 30019, South Korea.}

\twocolumn[
\begin{@twocolumnfalse}
\maketitle
%&&&&&&&&&&&&&&&<ABSTRACT>&&&&&&&&&&&&&&
\begin{abstract}
\section*{Abstract}

Through ordinary transmissions over wireless multicast networks are greatly hampered due to the simultaneous presence of fading and shadowing of wireless channels, secure transmissions can be enhanced by properly exploiting random attributes of the propagation medium. This study focuses on the utilization of those attributes to enhance the physical layer security (PLS) performance of a dual-hop wireless multicast network over $\kappa-\mu$ shadow-fading channel under the wiretapping attempts of multiple eavesdroppers. In order to improve the secrecy level, the best relay selection strategy among multiple relays is employed. Performance analysis is carried out based on the mathematical modeling in terms of analytical expressions of non-zero secrecy capacity probability, secure outage probability, and ergodic secrecy capacity over multicast relay networks. Capitalizing on those expressions, the effects of system parameters, i.e., fading, shadowing, the number of antennas, destination receivers, eavesdroppers, and relays, on the secrecy performance are investigated. Numerical results show that the detrimental impacts caused by fading and shadowing can be remarkably mitigated using the well-known opportunistic relaying technique. Moreover, the proposed model unifies secrecy analysis of several classical models, thereby exhibiting enormous versatility than the existing works.
\end{abstract}
%&&&&&&&&&&&&&<END-ABSTRACT>&&&&&&&&&&&&
%$$$$$$$$$$$$$$$<KEYWORDs>$$$$$$$$$$$$$$
\begin{IEEEkeywords}
\section*{Keywords} 

$\kappa-\mu$ shadowed fading, opportunistic relaying, physical layer security, secure outage probability, wireless multicasting.

\end{IEEEkeywords}
%$$$$$$$$$$$$$<END-KEYWORD>$$$$$$$$$$$$$
\end{@twocolumnfalse}
]

%%%%%%%%%%%%%%%%<SECTION>%%%%%%%%%%%%%%%
\section{Introduction}

%#############<SUB-SECTION>#############
\subsection{Background and Related Works}
Data-carrying radio waves, which propagate through communication channels, experience several constraints, such as diffraction, scattering of waves on the object surface, shadowing, fading, limited bandwidth, and vulnerable nature of the wireless medium, etc. Especially, random shadowing caused by obstacles in the local scenarios or human body exhibits some variations in the interaction pattern of radio wave propagation. Therefore, the aim to build a fortified network not only involves the enhancement of communicating link performance but also the use of optimized protocols in transmitter and receiver circuitry to prevent shadowing and swift variations in multipath propagation conditions.  
%%%%%%%%%%%%%%%%%%%%%%%%%%%%%%% Group on Shadowing %%%%%%%%%%%%%%%%%%%%%%%%%%%%%%%%%%%%%%
To understand the effect of dominant and scattered components of the dual shadowing process, the authors in \cite{simmons2018double} derived expressions of probability density function (PDF), cumulative distribution function (CDF), and Moment Generating Function (MGF) of Rician-fading \textcolor{black}{envelopes}. In \cite{ermolova2012outage}, the authors analyzed outage performance over compound $\eta-\mu$ fading-log-normal shadowing radio channels and derived a formula for the PDF of $\eta-\mu$ fading distribution. With a view to unifying all classic fading models, the authors of \cite{zhang2015effective} investigated the ergodic capacity (EC) and flexibility of channels. Similarly, a trade-off between mathematical complexity and flexibility was represented in \cite{lopez2017kappa} by varying different fading parameters under Rayleigh fading distribution. An optimal rate adaptation (ORA) scheme under composite $\kappa-\mu$ / Inverse Gamma (I-Gamma) and $\eta-\mu$ / I-Gamma \cite{yoo2017kappa} fading models was investigated in \cite{pant2020channel}, where the analysis incorporated expressions of channel capacity (CP) with a view to developing highly attractive wireless communication systems. The authors in \cite{pant2019error} studied average symbol error probability (SEP), the CP under ORA, channel inversion with fixed-rate (CIFR), and truncated CIFR under I-Gamma shadowed fading channels. A qualified analysis among Log-Normal, Inverse \textcolor{black}{Gaussian} \cite{raghuwanshi2021alpha}, Gamma, and I-Gamma distribution was also depicted. Apart from these fading channels, the $\kappa-\mu$ shadowed fading channel gained much popularity because of its broad spectrum of flexibility and general characterization.  The researchers in \cite{chun2017comprehensive} proposed a $\kappa-\mu$ shadowed fading model to improve network quality, capacity, and spectral efficiency taking human body shadowing under consideration. Maximum ratio combining (MRC) and square-law combining schemes over $\kappa-\mu$ shadowed fading channel was performed in \cite{al2016analysis} to observe energy detection in wireless communication scenarios. {\color{black} The $\kappa-\mu$ shadowed model was also employed in \cite{aalo2020impact} to investigate the bit error rate (BER) considering user mobility instead of static user position. The outage probability (OP), average bit error probability and the effective capacity was analyzed in \cite{al2021unified} considering double shadowed $\kappa-\mu$ fading model.} With the interest of investigating the effective rate of the multiple-input multiple-output (MIMO) systems, the authors in \cite{li2017effective,zhang2016high} performed higher-order statistics analysis and proved that the properties of the considered fading distribution could be approximated by Gamma distribution. Multiplicative shadowing was investigated in {\color{black} many works such as} \cite{yoo2016shadowed, subhash2019asymptotic}, where the authors showed the superiority of $\kappa-\mu$ / Gamma composite fading model over $\kappa-\mu$ / log-normal \textcolor{black}{line-of-sight} (LOS) shadowed fading model in an indoor off-body communication system. 

%%%%%%%%%%%%% Group on Shadowing + Security + Multiple Eavesdroppers %%%%%%%%%%%%%%%%%%%%%%%

Presently, the need for security enhancement between communicating devices in wireless medium has become a major concern \cite{badrudduza2021secrecy}. To compensate for the consequences due to several hindrances such as shadowing, fading, wiretapping, various studies have been introduced to build reliable wireless networks to make it impossible for any eavesdropper to decode any information from the communicating network. Wyner’s classic wiretap structure is known as the leading model which recapitulates the importance of security enhancement over different fading channels. Among them, shadow-fading distribution is more admissible for being amenable than any other modern fading model and vast span of propagation conditions. In an extension of this fact, outage performance over log-normal shadowed Rayleigh fading channel was analyzed \cite{han2016outage, zhang2017secrecy}, where the authors prosecuted the \textcolor{black}{Gaussian}-Hermite integration approach to show that outage performance significantly enhances standard deviations of shadowing. Authors in \cite{vegasanchez2021information} analyzed the impact of shadowing on numbers of antennas and propagation conditions over secrecy performance deriving several secrecy measures. Free space optical (FSO) links undergoing shadowed Rician and $\alpha-\mu$ fading in \cite{sumona2021security} were analyzed to achieve a perfect secrecy level despite severe channel constraints. To get the more generalized picture, security over $\kappa-\mu$ shadowed fading model at a physical layer was examined in \cite{jiang2020physical, nunes2017physical, ai2019secrecy}, where the authors demonstrated that MIMO system diversity manifests superior performance over MRC and selecting diversity in case of security. The authors in \cite{sun2019secrecy} considered $\kappa-\mu$ shadow-fading channels to observe the effect of correlation and drew a conclusion on the fact that a correlation coefficient works as a propitious parameter in case of improving secrecy performance at  a lower signal-to-noise ratio (SNR).

%%%%%%%%%%%%%%%%%%%% Group on Shadowing + Opportunistic Relaying %%%%%%%%%%%%%%%%%%%%%%%%%%%

Cooperative relaying is another proficient technology to enhance security in wireless systems \cite{badrudduza2020enhancing}. The impact of shadowing is inevitably minimized using the dual-hop network which in turn helps to improve wireless links. {\color{black}The authors in \cite{kalantari2011performance} analyzed the OP over a non-identical log normal fading channel employing the "best relay" selection scheme.} A multi-branch-multi-hop cooperative relaying system in the presence of co-channel interferers was considered in \cite{yu2012general,feng2017performance} over shadowed Nakagami-$m$ channel, where the authors evinced that a fading parameter affects the system performance greater than the shadowing parameter. A multiple cooperative relay-based satellite-terrestrial system over non-identical shadowed Rician and Nakagami-$m$ fading channels was studied in \cite{iqbal2013integrated, fan2017secure}, whether the authors came to a conclusion that EC of the system decreases with the number of relay nodes between satellite and ground users. The best relay selection scheme was employed in the performance analysis of $\kappa-\mu$ shadowed fading channel with multiple relays in \cite{zhang2017performance}. The authors examined the expressions of outage probability (OP), EC, and average BER to manifest the superiority of multiple relay systems over all other transmission techniques. The performance of a $\kappa-\mu$ shadowed fading model with beamforming and amplify-and-forward (AF) relaying technique was demonstrated in \cite{arti2016beamforming, zou2016relay}. The author validated the fact that the increment in the number of antennas at the transmitting earth station (ES) provides better performance than the increment in the number of antennas at the receiving ES.

%%%%%%%%%%%%%%%%%%%%%%%%%%%%%%%%%%%%% Group on Multicasting%%%%%%%%%%%%%%%%%%%%%%%%%%%%%%%
Recently, multicast channels have earned much popularity in wireless communication due to its versatile nature to transmit data to multiple destinations accommodating fewer network supplies \cite{kundu2019enhancing}. Due to such wide spread of wireless multicast networks, lots of researches have been are undertaken to enhance the secrecy performance. Physical layer security (PLS) in a multicasting scenario was analyzed in \cite{shrestha2013secure} over quasi-static Rayleigh fading channel where the authors derived the expressions of the probability of non-zero secrecy multicast capacity (PNSMC) and secure outage probability for multicasting (SOPM). A virtual MIMO antenna array scheme was employed in a multiple relay system in \cite{wang2014outage}, where the authors provided a completed description on cooperative spatial multiplexing and derived analytical expressions of SOPM and ergodic secrecy multicast capacity (ESMC).  The authors proved that the secrecy performance of a cooperative network completely outweighs that of a direct network.

%#############<SUB-SECTION>#############
\subsection{Motivation and Contributions}

According to the aforementioned works, most of the researches was performed to investigate the system improvement applying numerous technologies over various fading channels, i.e., both generalized and multipath fading. However, there are few works that considered wireless multicasting scenarios and the impact of shadowing on the secure wireless multicast schemes with opportunistic relaying and multiple eavesdroppers over multipath/generalized shadowed fading channels has not been investigated yet. Motivated from this perspective, this paper investigates the mathematical modeling of a secure wireless multicasting scheme over $\kappa-\mu$ shadow-fading channels with an opportunistic relaying technique. Here, a single sender communicates with a set of multiple destination receivers via a set of multiple cooperative relays in the existence of multiple eavesdroppers. $\kappa-\mu$ shadow model assumes random fluctuations of ling of sight (LOS) component and also matches well with experimental data of land mobile satellite (LMS) communication channel. Moreover, a $\kappa-\mu$ fading channel is a generalized fading model, and hence a number of classical fading models, e.g., one-sided \textcolor{black}{Gaussian}, Nakagami-$m$, Rayleigh, Rician-$K$, and shadowed Rician, can be obtained as particular cases of the proposed model. The prime contributions of the authors are as follows:
\begin{itemize} 

\item We derive the PDF and CDF of SNRs for multicast and eavesdropper channels by first realizing the PDF of SNRs of each individual hop and then obtaining the PDF of dual-hop SNR with the best relay selection algorithm. To the best of the authors’ knowledge, the derived PDF and CDF are absolutely novel and have not been reported yet in any existing literature.

\item We analyze the secrecy performance  utilizing novel expressions of some well-known secrecy metrics i.e. PNSMC, SOPM, and ESMC, and quantify the effects of each system parameter, i.e., fading parameters, shadowing, number of receive antennas, relays, destination receivers, and eavesdroppers, etc. In comparison to the previous literature, only the proposed work demonstrates how the detrimental impact of shadowing on secure multicasting can be mitigated {by} employing an opportunistic relaying strategy.

\item Finally, we verify all the numerical results corresponding to the derived analytical expressions of secrecy metrics via Monte-Carlo simulations.

\end{itemize}

\subsection{Organization}
The organization of this paper is summarized as follows: Section \ref{s1} discusses the proposed system model and problem formulation. The derivations of the expressions of secrecy metrics i.e. PNSMC, SOPM, and ESMC are demonstrated in Section \ref{s2}. Section \ref{s3} provides the numerical results analysis. Finally, the conclusion of this work is illustrated in section \ref{s4}.

%%%%%%%%%%%%%%%%<SECTION>%%%%%%%%%%%%%%%
\section{System Model and Problem Formulation}
\label{s1}
%===============<FIGURE>================
\begin{figure*}[!ht]
\vspace{0mm}
   \centerline{\includegraphics[width=0.7\textwidth]{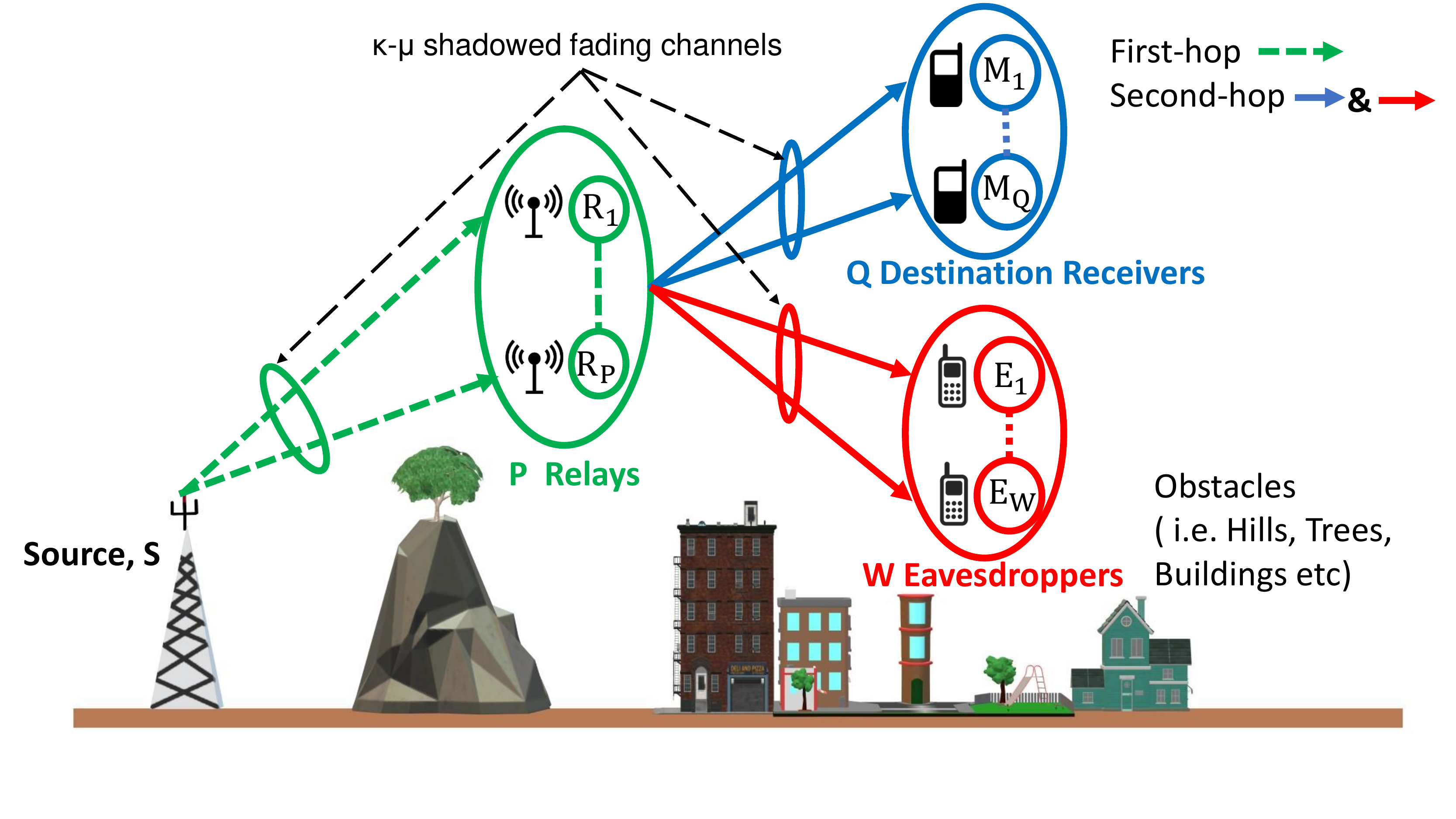}}
        \vspace{0mm}
    \caption{Proposed system model.}
    \label{fig.1}
\end{figure*}
%=============<END-FIGURE>==============
A secure wireless multicast network is shown in Fig.\ref{fig.1}, where a source, $S$ with a single antenna sends secret information to a set of $Q$ destination receivers via $P$ relays. A set of $W$ eavesdroppers are also present in that network which is intended to decipher the secret messages. Each relay has a single antenna while each destination receiver and each eavesdropper are equipped with $G_{Q}$ and $G_{W}$ antennas, respectively. In this particular communication scenario, we assume that the distances of $Q$ and $W$ from $S$ are too large and due to masking effect and severe shadowing there are no direct communication paths between $S$ to $Q$ as well as $S$ to $W$. Hence the only communication path that exists is through the relay. The overall process is performed in two phases. In the first-hop, $S$ sends messages to the relay. Then, in the second-hop, desired messages are received by the destination receivers from the best relay only. {\color{black} It is noteworthy that AF variable gain relaying scheme has been adopted in this work.} Meanwhile, at the same time slot, the eavesdroppers also try to steal information from a best relay.

{\color{black}The best relay selection is performed using the method of distributed timers, where all the relays use their timers to estimate own instantaneous channel gains and compete to access the wireless medium according to their own channel conditions. In an opportunistic relaying scheme, competition among cooperative relays offers diversity benefits in the direction of destination that enhances secrecy rate (i.e. minimises secure outage probability) and adhere to the ‘opportunistic’ cooperation rule giving priority to the ‘best’ available relay even when they are not chosen to transmit but rather chosen to cooperatively listen.}

The direct channel coefficients for $S$ to $a$th $(a=1,2,3,\ldots,P)$ relay link is $f_{s,a}\in \mathcal{C}^{1\times1}$, for $a$th relay to $b$th $(b=1,2,3,\ldots,Q)$ destination receiver link is $\mathbf{g}_{a,b}\in \mathcal{C}^{G_{Q}\times1}$ $(\text{i.e. }\mathbf{g}_{ab}=$ $[g_{1ab}\quad$ $g_{2ab}\quad$
$g_{3ab}\quad$ $\ldots\quad g_{G_{Q}ab}]^{T})$ and for $a$th relay to $c$th $(c=1,2,3,\ldots,W)$ eavesdropper link is $\mathbf{h}_{a,c}\in \mathcal{C}^{G_{W}\times1}$ $(\text{i.e. }\mathbf{h}_{ac}=$ $ [h_{1ac}\quad$  $h_{2ac}\quad$ $h_{3ac}\quad$ $\ldots\quad h_{G_{W}ac}]^{T})$.

In the first-hop, the received signal at $a$th relay is expressed as
\begin{align}
\label{a1}
y_{s,a}=f_{s,a}x+z_{a},
\end{align}
where $x\sim \mathcal{\widetilde{N}}(0, T_{s})$ is the transmitted message signal from $S$, $T_{s}$ is the transmit power, \textcolor{black}{$\mathcal{\widetilde{N}}$ is   circularly symmetric complex Gaussian distribution
with mean 0 and variance $T_{s}$}, $z_{a}\sim \mathcal{\widetilde{N}}(0, N_{a})$ indicates the additive white gaussian noise (AWGN) imposed on $a$th relay with noise power $N_{a}$.

In the second-hop, the received signal of the $a$th best relay will be forwarded to the receivers. Hence, the received signals at the $b$th receiver is denoted as
\begin{align}
\nonumber
        \mathbf{y}_{a,b}&=\mathbf{g}_{a,b}y_{s,a}+\mathbf{j}_{b}=\mathbf{g}_{a,b}(f_{s,a}x+z_{a})+\mathbf{j}_{b}
\\
\label{a2}
         &=\mathbf{d}_{ab}x+\mathbf{u}_{b},
\end{align}
whereas the received signal at $c$th eavesdropper can be written as 
\begin{align}
\nonumber
        \mathbf{y}_{a,c}&=\mathbf{h}_{a,c}y_{s,a}+\mathbf{k}_{c}=\mathbf{h}_{ac}(f_{s,a}x+z_{a})+\mathbf{k}_{c}
\\
\label{a3}
         &=\mathbf{d}_{ac}x+\mathbf{v}_{c}.
\end{align}
Here, $\mathbf{u}_{b} \triangleq \mathbf{g}_{a,b}z_{a}+\mathbf{j}_{b}$, $\mathbf{v}_{c} \triangleq \mathbf{h}_{a,c}z_{a}+\mathbf{k}_{c}$, $\mathbf{d}_{a,b}= \mathbf{g}_{a,b}f_{s,a}$, $\mathbf{d}_{a,c}= \mathbf{h}_{a,c}f_{s,a}$, $\mathbf{j}_{b}\sim \mathcal{\widetilde{N}}(0, N_{b}\mathbf{I}_{G_{Q}})$ and $\mathbf{k}_{c}\sim \mathcal{\widetilde{N}}(0, N_{c}\mathbf{I}_{G_{W}})$ symbolize the noises imposed on the $b$th receiver and $c$th eavesdropper, $N_{b}$ and $N_{c}$ represent noise powers, and $\mathbf{I}_{G}(.)$ is the identity matrix of order $G\times G$.

We assume all the channels between source to relays ($S\rightarrow P$ links), relays to destination receivers ($P\rightarrow Q$ links), and relays to eavesdroppers ($P\rightarrow W$ links) undergo independent and identically distributed (i.i.d.) $\kappa-\mu$ shadowed fading i.e. all the LOS components are subjected to shadowing.  This channel is widely used to model land mobile satellite communication systems. Moreover, this model exhibits extreme versatility since a wide number of multipath/generalized fading models can be obtained as special cases from this particular model as shown in Table \ref{table:1}.
%*************<END-EQUATION>************

%%%%%%%%%%%%%%%%%%%%% Table 1 %%%%%%%%%%%%%%% 
\begin{table}[h!]
\centering
\caption{Special Cases of $\kappa-\mu$ shadowed fading channel \cite{paris2013statistical}.}
\label{table:1}
\begin{tabular}{ m{10em}|c|c|c } 
\hline
\multicolumn{1}{c|}{\multirow{2}{*}{Fading Channels}} & \multicolumn{3}{l}{$\kappa-\mu$ Shadowed Fading Parameters} 
\\
\multicolumn{1}{c|}{}                   & $\kappa_{b}$=$\kappa_{c}$=$\kappa_{a}$    & $\mu_{b}$=$\mu_{c}$=$\mu_{a}$     & $m_{b}$=$m_{c}$=$m_{a}$  
\\ \hline \hline
One Sided Gaussian & 0 & 0.5 & $\infty$
\\\hline
Rayleigh & 0 & 1 & $\infty$
\\ \hline
Nakagami-$m$ & 0 & $m$ & $\infty$
\\\hline
Shadowed Rician & $K$ & 1 & $m$
\\\hline
Rician-$K$ & $K$ & 1 & $\infty$ 
\\\hline
\end{tabular}
\end{table}
%<<<<<<<<<<<<<<<END-TABLE>>>>>>>>>>>>>>>
\subsection{Channel Model}
The instantaneous SNRs  of $S \rightarrow P$, $P \rightarrow Q$ and $P \rightarrow W$ links are respectively given by $\lambda_{s,a}=\frac{T_{s}}{N_{a}}\|f_{s,a}\|^{2}$, $\lambda_{a,b}=\frac{P_{a}}{N_{b}}\|\mathbf{g}_{a,b}\|^{2}$, and $\lambda_{a,c}=\frac{P_{a}}{N_{c}}\|\mathbf{h}_{a,c}\|^{2}$, where $P_{a}$ is the transmit signal power from the relay. The PDF of respective SNRs are shown below.

\subsubsection{PDF of $\lambda_{s,a}$}
The PDF of $\lambda_{s,a}$ is given by  \cite[eq.~1]{srinivasan2018secrecy}
\setcounter{eqnback}{\value{equation}} \setcounter{equation}{3}
\begin{align}
\label{a5}
f_{s,a}(\lambda)=\alpha_{1}e^{-\mathcal{A}_{2}\lambda}\lambda^{\mu_{a}-1}{_{1}F_{1}}(m_{a}, \mu_{a}; \alpha_{3}\lambda),
\end{align}
where $\alpha_{1}=\frac{\mu_{a}^{\mu_{a}}m_{a}^{m_{a}}(1+\kappa_{a})^{\mu_{a}}}{\Gamma(\mu_{a})\bar{\lambda}_{sa}^{\mu_{a}} (\mu_{a}\kappa_{a}+m_{a})^{m_{a}}}$,
$\mathcal{A}_{2}=\frac{\mu_{a}(1+\kappa_{a})}{\bar{\lambda}_{sa}}$,
\\
$\alpha_{3}=\frac{\mu_{a}^{2}\kappa_{a}(1+\kappa_{a})}{(\mu_{a}\kappa_{a}+m_{a})\bar{\lambda}_{sa}}$,
the average SNR of $S\rightarrow P$ link channel is $\bar{\lambda}_{sa}$, $\kappa_{a}$ is the ratio of the powers between dominant and scattered components, $\mu_{a}$ is the number of clusters, $m_{a}$ is the Nakagami-$m$ faded shadowing component and
$_{1}F_{1}(.,.;.)$ is the confluent hyper-geometric function which can be expressed as $_{1}F_{1}(x_{1},y_{1};z_{1})=\frac{\Gamma(y_{1})}{\Gamma(x_{1})}\sum_{d_{1}=0}^{\infty}\frac{\Gamma(x_{1}+d_{1})z_{1}^{d_{1}}}{\Gamma(y_{1}+d_{1})d_{1}!}$ \cite[eq.~13]{al2017multiantenna}. Hence, finally $f_{s,a}(\lambda)$ can be written as
\begin{align}
\label{a6}
f_{s,a}(\lambda)=\sum_{e_{1}=0}^{\infty}\mathcal{A}_{1}e^{-\mathcal{A}_{2}\lambda}\lambda^{\mathcal{A}_{3}},
\end{align}
where $\mathcal{A}_{1}=\alpha_{1}\alpha_{e_{1}}$, $\alpha_{e_{1}}=\frac{\Gamma(\mu_{a})\Gamma(m_{a}+e_{1})\alpha_{3}^{e_{1}}}{\Gamma(m_{a})\Gamma(\mu_{a}+e_{1})e_{1}!}$, and $\mathcal{A}_{3}=\mu_{a}-1+e_{1}$. 

%*************************************************************************************************
\subsubsection{PDF of $\lambda_{a,b}$}
Similar to \eqref{a6}, PDF of $\lambda_{a,b}$ can be written as  \cite[eq.~1]{srinivasan2018secrecy}
\begin{align}
\label{a7}
f_{a,b}(\lambda)=\sum_{e_{2}=0}^{\infty}\mathcal{B}_{1}e^{-\mathcal{B}_{2}\lambda}\lambda^{\mathcal{B}_{3}},
\end{align}
where $\mathcal{B}_{1}=\beta_{1}\beta_{e_{2}}$,$\beta_{1}=\frac{(G_{Q}\mu_{b})^{G_{Q}\mu_{b}}(G_{Q}m_{b})^{G_{Q}m_{b}}(1+\kappa_{b})^{G_{Q}\mu_{b}}}{\Gamma(G_{Q}\mu_{b})(\bar{\lambda}_{ab})^{G_{Q}\mu_{b}} (G_{Q}\mu_{b}\kappa_{b}+G_{Q}m_{b})^{G_{Q}m_{b}}}$, $\beta_{e_{2}}=\frac{\Gamma(G_{Q}\mu_{b})\Gamma(G_{Q}m_{b}+e_{2})\beta_{2}^{e_{2}}}{\Gamma(G_{Q}m_{b})\Gamma(G_{Q}\mu_{b}+e_{2})e_{2}!}$,$\beta_{2}=\frac{G_{Q}^{2}\mu_{b}^{2}\kappa_{b}(1+\kappa_{b})}{(G_{Q}\mu_{b}\kappa_{b}+G_{Q}m_{b})\bar{\lambda}_{ab}}$, $\mathcal{B}_{2}=\frac{G_{Q}\mu_{b}(1+\kappa_{b})}{\bar{\lambda}_{ab}}$, $\mathcal{B}_{3}=G_{Q}\mu_{b}-1+e_{2}$, the average SNR of $P\rightarrow Q$ link channel is $\bar{\lambda}_{ab}$ and shape parameters corresponding to $P\rightarrow Q$ link are denoted by $\kappa_{b}$, $\mu_{b}$ and $m_{b}$.

%*************************************************************************************************
\subsubsection{PDF of $\lambda_{a,c}$}
The PDF of $\lambda_{a,c}$ can be expressed as \cite[eq.~1]{srinivasan2018secrecy} 
\begin{align}
\label{a8}
f_{a,c}(\lambda)=\sum_{e_{3}=0}^{\infty} \mathcal{C}_{1}e^{-\mathcal{C}_{2}\lambda}\lambda^{\mathcal{C}_{3}},
\end{align}
where $\mathcal{C}_{1}=\iota_{1}\iota_{e_{3}}$,$\iota_{1}=\frac{(G_{W}\mu_{c})^{G_{W}\mu_{c}}(G_{W}m_{c})^{G_{W}m_{c}}(1+\kappa_{c})^{G_{W}\mu_{c}}}{\Gamma(G_{W}\mu_{c})(\bar{\lambda}_{ac})^{G_{W}\mu_{c}} (G_{W}\mu_{c}\kappa_{c}+G_{W}m_{c})^{G_{W}m_{c}}}$, $\iota_{e_{3}}=\frac{\Gamma(G_{W}\mu_{c})\Gamma(G_{W}m_{c}+e_{3})\iota_{2}^{e_{3}}}{\Gamma(G_{W}m_{c})\Gamma(G_{W}u_{c}+e_{3})e_{3}!}$, $\iota_{2}=\frac{G_{W}^{2}\mu_{c}^{2}\kappa_{c}(1+\kappa_{c})}{(G_{W}\mu_{c}\kappa_{c}+G_{W}m_{c})\bar{\lambda}_{ac}}$,
$\mathcal{C}_{2}=\frac{G_{W}\mu_{c}(1+\kappa_{c})}{\bar{\lambda}_{ac}}$,
$\mathcal{C}_{3}=G_{W}\mu_{c}-1+e_{3}$,
the average SNR of $P\rightarrow W$ link is $\bar{\lambda}_{ac}$ and $\kappa_{c}$, $\mu_{c}$ and $m_{c}$ symbolize the shape parameters corresponding to $P\rightarrow W$ link.

\subsection{PDFs of Dual-hop SNRs}
Denoting SNRs of $S\rightarrow Q$ and $S\rightarrow W$ links by $\lambda_{s,b}$ and $\lambda_{s,c}$, respectively, the PDFs of $\lambda_{s,b}$ and $\lambda_{s,c}$ are defined as
\begin{align}
\label{a9}
f_{s,b}(\lambda)=\frac{dF_{s,b}(\lambda)}{d\lambda},
\\
\label{a10}
f_{s,c}(\lambda)=\frac{dF_{s,c}(\lambda)}{d\lambda},
\end{align} 
where $F_{s,b}(\lambda)$, and $F_{s,c}(\lambda)$ express the CDFs of $\lambda_{s,b}$, and $\lambda_{s,c}$. The CDF of $\lambda_{s,b}$ is defined as 
 \begin{align}
 \label{a11}
F_{s,b}(\lambda)=1-Pr(\lambda_{s,a}>\lambda_{s,b})Pr(\lambda_{a,b}>\lambda_{s,b}),
\end{align}
where $Pr(\lambda_{s,a}>\lambda_{s,b})$ and $Pr(\lambda_{a,b}>\lambda_{s,b})$ are the complementary cumulative distribution functions (CCDFs) of $\lambda_{s,a}$ and $\lambda_{a,b}$, and the CCDFs are respectively defined as 
\begin{align}
\label{a12}
Pr(\lambda_{s,a}>\lambda_{s,b})=\int_{\lambda_{s,b}}^{\infty}f_{s,a}(\lambda)d\lambda,
\\
\label{a13}
Pr(\lambda_{a,b}>\lambda_{s,b})=\int_{\lambda_{s,b}}^{\infty}f_{a,b}(\lambda)d\lambda.
\end{align}
Substituting \eqref{a6} into \eqref{a12} and executing integration using the following identity of \cite[eq.~3.351.2]{GR:07:Book}, we get
\begin{align}
\label{a14}
Pr(\lambda_{s,a}>\lambda_{s,b})=\sum_{e_{1}=0}^{\infty}\mathcal{A}_{1}\mathcal{A}_{2}^{-\mu_{a}-e_{1}}\Gamma(\mu_{a}+e_{1},\mathcal{A}_{2}\lambda_{s,b}),
\end{align}
\textcolor{black}{where $\Gamma(.,.)$ denotes the upper incomplete gamma function.}
Further, substituting \eqref{a7} into \eqref{a13} and performing integration, we have
\begin{align}
\label{a15}
Pr(\lambda_{a,b}>\lambda_{s,b})=\sum_{e_{2}=0}^{\infty}\mathcal{B}_{1}\mathcal{B}_{2}^{-G_{Q}\mu_{b}-e_{2}}\Gamma(G_{Q}\mu_{b}+e_{2},\mathcal{B}_{2}\lambda_{s,b}).
\end{align}
Deploying \eqref{a14} and \eqref{a15} into \eqref{a11}, the CDF of $\lambda_{s,b}$ is obtained as
\begin{align}
\nonumber
F_{s,b}(\lambda)&=1-\sum_{e_{2}=0}^{\infty}\sum_{e_{1}=0}^{\infty}\lambda_{e_{2}}\Gamma(\mu_{a}+e_{1},\mathcal{A}_{2}\lambda)
\\
\label{a16}
&\times\Gamma(G_{Q}\mu_{b}+e_{2},\mathcal{B}_{2}\lambda),
\end{align}
where $\lambda_{e_{2}}=\mathcal{A}_{1}\mathcal{B}_{1}\mathcal{A}_{2}^{-\mu_{a}-e_{1}}\mathcal{B}_{2}^{-G_{Q}\mu_{b}-e_{2}}$. Now, substituting \eqref{a16} into \eqref{a9} and performing differentiation with respect to $\lambda_{s,b}$, the PDF of $\lambda_{s,b}$ is found as
\begin{align}
\nonumber
f_{s,b}(\lambda)&=\sum_{s_{2}=0}^{\infty}\sum_{e_{4}=0}^{\infty}\frac{\lambda_{e_{4}}\lambda^{\mu_{a}+s_{2}-1}}{e^{\mathcal{A}_{2}\lambda}}\Gamma(G_{Q}\mu_{b}+e_{4},\mathcal{B}_{2}\lambda)
\\
\label{a17}
&+\sum_{s_{3}=0}^{\infty}\sum_{e_{5}=0}^{\infty}\frac{\lambda_{e_{5}}\lambda^{G_{Q}\mu_{b}+e_{5}-1}}{e^{\mathcal{B}_{2}\lambda}}\Gamma(\mu_{a}+s_{3},\mathcal{A}_{2}\lambda),
\end{align}
where $\lambda_{e_{4}}=\mathcal{A}_{1}\mathcal{B}_{1}\mathcal{A}_{2}^{-\mu_{a}-s_{2}}\mathcal{B}_{2}^{-G_{Q}\mu_{b}-e_{4}}$
and 
\\
$\lambda_{e_{5}}=\mathcal{A}_{1}\mathcal{B}_{1}\mathcal{A}_{2}^{-\mu_{a}-s_{3}}\mathcal{B}_{2}^{-G_{Q}\mu_{b}-e_{5}}$.
Similarly, the CDF of $\lambda_{s,c}$ can be obtained as
\begin{align}
\nonumber
F_{s,c}(\lambda)&=1-\sum_{e_{3}=0}^{\infty}\sum_{e_{1}=0}^{\infty}\lambda_{e_{3}}\Gamma(\mu_{a}+e_{1},\mathcal{A}_{2}\lambda)
\\
\label{a18}
&\times \Gamma(G_{W}\mu_{c}+e_{3},\mathcal{C}_{2}\lambda),
\end{align}
where
$\lambda_{e_{3}}=\mathcal{A}_{1}\mathcal{C}_{1}\mathcal{A}_{2}^{-\mu_{a}-e_{1}}\mathcal{C}_{2}^{-G_{W}\mu_{c}-e_{3}}$. Furthermore, replacing \eqref{a18} into \eqref{a10}, the PDF of $\lambda_{s,c}$ is obtained as
\begin{align}
\nonumber
f_{s,c}(\lambda)&=\sum_{s_{4}=0}^{\infty}\sum_{e_{6}=0}^{\infty}\frac{\lambda_{e_{6}}\lambda^{\mu_{a}+s_{4}-1}}{e^{\mathcal{A}_{2}\lambda}}\Gamma(G_{W}\mu_{c}+e_{6},\mathcal{C}_{2}\lambda)
\\
\label{a19}
&+\sum_{s_{5}=0}^{\infty}\sum_{e_{7}=0}^{\infty}\frac{\lambda_{e_{7}}\lambda^{G_{W}\mu_{c}+e_{7}-1}}{e^{\mathcal{C}_{2}\lambda}}\Gamma(\mu_{a}+s_{5},\mathcal{A}_{2}\lambda),
\end{align}
where $\lambda_{e_{6}}=\mathcal{A}_{1}\mathcal{C}_{1}\mathcal{A}_{2}^{-\mu_{a}-s_{4}}\mathcal{C}_{2}^{-G_{W}\mu_{c}-e_{6}}$
and 
\\
$\lambda_{e_{7}}=\mathcal{A}_{1}\mathcal{C}_{1}\mathcal{A}_{2}^{-\mu_{a}-s_{5}}\mathcal{C}_{2}^{-G_{W}\mu_{c}-e_{7}}$.

%*************************************************************************************************
\subsection{Best Relay Selection}
Let $\lambda_{b}^{*}$ denote the SNR between best relay and $b$th receiver which is expressed as
\begin{align}
\label{a20}
\lambda_{b}^{*}= arg_{a\in\varpi}^{max} min(\lambda_{s,a},\lambda_{a,b}),
\end{align}
where $\varpi={1,2,\ldots,P}$ is the relay set. The CDF of $\lambda_{b}^{*}$ is demonstrated as
\begin{align}
\label{a21}
F_{*,b}(\lambda)= [F_{s,b}(\lambda)]^{P}.
\end{align}
Hence, substituting \eqref{a16} into \eqref{a21}, the CDF of $\lambda_{b}^{*}$ is derived as
\begin{align}
\nonumber
F_{*,b}(\lambda)&= \Big[1-\sum_{e_{2}=0}^{\infty}\sum_{e_{1}=0}^{\infty}\lambda_{e_{2}}\Gamma(\mu_{a}+e_{1},\mathcal{A}_{2}\lambda)
\\
\label{a22}
&\times\Gamma(G_{Q}\mu_{b}+e_{2},\mathcal{B}_{2}\lambda)\Big]^{P}.
\end{align}
Differentiating \eqref{a21} with respect to $\lambda_{s,b}$, the PDF of $\lambda_{b}^{*}$ is obtained as
\begin{align}
\label{a23}
f_{*,b}(\lambda)=Pf_{s,b}(\lambda)[F_{s,b}(\lambda)]^{P-1}.
\end{align}
Again, deploying \eqref{a16} and \eqref{a17} into \eqref{a23}, $f_{*,b}(\lambda)$ is given by
\begin{align}
\nonumber
& f_{*,b}(\lambda)
\\
\nonumber
&=P\biggl[\sum_{s_{2}=0}^{\infty}\sum_{e_{4}=0}^{\infty}\frac{\lambda_{e_{4}}\lambda^{\mu_{a}+s_{2}-1}}{e^{\mathcal{A}_{2}\lambda}}\Gamma(G_{Q}\mu_{b}+e_{4},\mathcal{B}_{2}\lambda)
\\
\nonumber
&+\sum_{s_{3}=0}^{\infty}\sum_{e_{5}=0}^{\infty}\frac{\lambda_{e_{5}}\lambda^{G_{Q}\mu_{b}+e_{5}-1}}{e^{\mathcal{B}_{2}\lambda}}\Gamma(\mu_{a}+s_{3},\mathcal{A}_{2}\lambda)\biggl]\Big[1
\\
\label{a24}
& -\sum_{e_{2}=0}^{\infty}\sum_{e_{1}=0}^{\infty}\lambda_{e_{2}}\Gamma(\mu_{a}+e_{1},\mathcal{A}_{2}\lambda)\Gamma(G_{Q}\mu_{b}+e_{2},\mathcal{B}_{2}\lambda)\Big]^{P-1}.
\end{align}
Similar to \eqref{a20}, the SNR between best relay and $c$th eavesdropper denoted by $\lambda_{c}^{*}$ is explained as 
\begin{align}
\label{a25}
\lambda_{c}^{*}= arg_{a\in\varpi}^{max} min(\lambda_{s,a},\lambda_{a,c}),
\end{align}
the CDF of which is given by
\begin{align}
\label{a26}
F_{*,c}(\lambda)= [F_{s,c}(\lambda)]^{P}.
\end{align}
Substituting \eqref{a18} into \eqref{a26}, the CDF of $\lambda_{c}^{*}$ is obtained as
\begin{align}
\nonumber
F_{*,c}(\lambda)&= \Big[1-\sum_{e_{3}=0}^{\infty}\sum_{e_{1}=0}^{\infty}\lambda_{e_{3}}\Gamma(\mu_{a}+e_{1},\mathcal{A}_{2}\lambda)
\\
\label{a27}
&\times\Gamma(G_{W}\mu_{c}+e_{3},\mathcal{C}_{2}\lambda)\Big]^{P}.
\end{align}
Further, differentiating \eqref{a26} with respect to $\lambda_{s,c}$, and substituting \eqref{a18} and \eqref{a19} into it, the PDF of $\lambda_{c}^{*}$ is derived as
\begin{align}
\nonumber
& f_{*,c}(\lambda)
\\
\nonumber
&=P\biggl[\sum_{s_{4}=0}^{\infty}\sum_{e_{6}=0}^{\infty}\frac{\lambda_{e_{6}}\lambda^{\mu_{a}+s_{4}-1}}{e^{\mathcal{A}_{2}\lambda}}\Gamma(G_{W}\mu_{c}+e_{6},\mathcal{C}_{2}\lambda)
\\
\nonumber
&+\sum_{s_{5}=0}^{\infty}\sum_{e_{7}=0}^{\infty}\frac{\lambda_{e_{7}}\lambda^{G_{W}\mu_{c}+e_{7}-1}}{e^{\mathcal{C}_{2}\lambda}}\Gamma(\mu_{a}+s_{5},\mathcal{A}_{2}\lambda)\biggl]\Big[1
\\
\label{a28}
& -\sum_{e_{3}=0}^{\infty}\sum_{e_{1}=0}^{\infty}\lambda_{e_{3}}\Gamma(\mu_{a}+e_{1},\mathcal{A}_{2}\lambda)\Gamma(G_{W}\mu_{c}+e_{3},\mathcal{C}_{2}\lambda)\Big]^{P-1}.
\end{align}

%*************************************************************************************************
\subsection{Modeling of Multicast Channels}
Note that we consider multiple destination receivers ($Q$) each of which can receive the multicast messages at the same instant. To ascertain a secure communication with each receiver, we demonstrate secrecy analysis considering the worst possible scenario which includes taking into consideration the minimum SNR among all receivers as denoted by $\lambda_{min}=min_{1<b<Q}\lambda_{b}^{*}$. Hence, it is clear from this consideration that if the proposed system is capable of protecting multicast information from being eavesdropped for the worst case, then for all other cases (i.e. better than worst case), the system will undoubtedly be secure. Since, $\lambda_{1}^{*}$, $\lambda_{2}^{*}$, $\ldots$, $\lambda_{Q}^{*}$ are all independent, using order statistics, the PDF of $\lambda_{min}$ can be defined as \cite{nafis2021secrecy}
\begin{align}
\label{a29}
f_{\lambda_{min}}(\lambda)=Q f_{*,b}(\lambda)[1-F_{*,b}(\lambda)]^{Q-1}.
\end{align}
Now, substituting \eqref{a22} and \eqref{a24} into \eqref{a29}, $f_{\lambda_{min}}(\lambda)$ is obtained as
\begin{align}
\nonumber
& f_{\lambda_{min}}(\lambda)
\\
\nonumber
&=PQ\biggl[\sum_{s_{2}=0}^{\infty}\sum_{e_{4}=0}^{\infty}\frac{\lambda_{e_{4}}\lambda^{\mu_{a}+s_{2}-1}}{e^{\mathcal{A}_{2}\lambda}}\Gamma(G_{Q}\mu_{b}+e_{4},\mathcal{B}_{2}\lambda)
\\
\nonumber
&+\sum_{s_{3}=0}^{\infty}\sum_{e_{5}=0}^{\infty}\frac{\lambda_{e_{5}}\lambda^{G_{Q}\mu_{b}+e_{5}-1}}{e^{\mathcal{B}_{2}\lambda}}\Gamma(\mu_{a}+s_{3},\mathcal{A}_{2}\lambda)\biggl]\Big[1
\\
\nonumber
& -\sum_{e_{2}=0}^{\infty}\sum_{e_{1}=0}^{\infty}\lambda_{e_{2}}\Gamma(\mu_{a}+e_{1},\mathcal{A}_{2}\lambda)\Gamma(G_{Q}\mu_{b}+e_{2},\mathcal{B}_{2}\lambda)\Big]^{P-1}
\\
\nonumber
&\times \Big[1-\big[1-\sum_{e_{2}=0}^{\infty}\sum_{e_{1}=0}^{\infty}\lambda_{e_{2}}\Gamma(\mu_{a}+e_{1},\mathcal{A}_{2}\lambda)
\\
\label{a30}
&\times \Gamma(G_{Q}\mu_{b}+e_{2},\mathcal{B}_{2}\lambda)\big]^{P}\Big]^{Q-1}.
\end{align}
Utilizing the identity of \cite[eq.~1.111]{GR:07:Book}, \eqref{a30} can be simplified as
\begin{align}
\nonumber
& f_{\lambda_{min}}(\lambda)
\\
\nonumber
&=PQ\sum_{e_{8}=0}^{Q-1}\sum_{e_{9}=0}^{P+Pe_{8}-1}\lambda_{e_{9}}\biggl[\sum_{s_{3}=0}^{\infty}\sum_{e_{5}=0}^{\infty}\frac{\lambda_{e_{5}}\lambda^{G_{Q}\mu_{b}+e_{5}-1}}{e^{\mathcal{B}_{2}\lambda}}
\\
\nonumber
&\times \Gamma(\mu_{a}+s_{3},\mathcal{A}_{2}\lambda)+\sum_{s_{2}=0}^{\infty}\sum_{e_{4}=0}^{\infty}\frac{\lambda_{e_{4}}\lambda^{\mu_{a}+s_{2}-1}}{e^{\mathcal{A}_{2}\lambda}}
\\
\nonumber
&\times \Gamma(G_{Q}\mu_{b}+e_{4},\mathcal{B}_{2}\lambda)\biggl]\Big[\sum_{e_{2}=0}^{\infty}\sum_{e_{1}=0}^{\infty}\lambda_{e_{2}}\Gamma(\mu_{a}+e_{1},\mathcal{A}_{2}\lambda)
\\
\label{a31}
&\times \Gamma(G_{Q}\mu_{b}+e_{2},\mathcal{B}_{2}\lambda)\Big]^{e_{9}},
\end{align}
where $\lambda_{e_{9}}= \frac{(_{\; e_{8}}^{Q-1})(_{\quad e_{9}}^{P+P e_{8}-1})}{(-1)^{-e_{8}-e_{9}}}$. Applying the identity of \cite[eq.~8.352.7]{GR:07:Book}, \eqref{a31} is further simplified as 
\begin{align}
\nonumber
f_{\lambda_{min}}(\lambda)&=PQ\sum_{e_{8}=0}^{Q-1}\sum_{e_{9}=0}^{P+Pe_{8}-1}\lambda_{e_{9}}e^{-(\mathcal{A}_{2}+\mathcal{B}_{2})\lambda}\big[\gamma_{1}(\lambda)\big]^{e_{9}}
\\
\nonumber
&\times \Big[\sum_{s_{3}=0}^{\infty}\sum_{e_{5}=0}^{\infty}\sum_{e_{11}=0}^{\mu_{a}+s_{3}-1}\lambda_{e_{11}}\lambda^{G_{Q}\mu_{b}+e_{5}+e_{11}-1} 
\\
\label{a32}
& +\sum_{s_{2}=0}^{\infty}\sum_{e_{4}=0}^{\infty}\sum_{e_{10}=0}^{G_{Q}\mu_{b}+e_{4}-1}\lambda_{e_{10}}\lambda^{\mu_{a}+s_{2}+e_{10}-1}\Big],
\end{align}
where $\lambda_{e_{10}}= \frac{\lambda_{e_{4}}\Gamma(e_{4}+G_{Q}\mu_{b})}{e_{10}!\mathcal{B}_{2}^{-e_{10}}}$ and
$\lambda_{e_{11}}= \frac{\lambda_{e_{5}}\Gamma(e_{5}+\mu_{a})}{e_{11}!\mathcal{A}_{2}^{-e_{11}}}$. Here $\gamma_{1}(\lambda)$ is denoted as 

\begin{align}
\nonumber
\gamma_{1}(\lambda)&=\sum_{e_{1}=0}^{\infty}\sum_{e_{2}=0}^{\infty}\sum_{e_{12}=0}^{\mu_{a}+e_{1}-1}\sum_{e_{13}=0}^{G_{Q}\mu_{b}+e_{2}-1} \lambda_{e_{13}}\lambda^{(e_{12}+e_{13})}
\\
\label{a33}
&\times e^{-(\mathcal{A}_{2}+\mathcal{B}_{2})\lambda},
\end{align}
where $\lambda_{e_{13}}=\frac{\mathcal{A}_{1}\mathcal{B}_{1}\Gamma(e_{1}+\mu_{a})\mathcal{B}_{2}^{-G_{Q}\mu_{b}-e_{2}+e_{13}}\Gamma(e_{2}+G_{Q}\mu_{b})
}{e_{12}!e_{13}!\mathcal{A}_{2}^{\mu_{a}+e_{1}-e_{12}}}$. Applying the multinomial theorem of \cite[eq.~7]{pena2014performance}, we obtain
\begin{align}
\nonumber
&[\gamma_{1}(\lambda)]^{e_{9}}
\\
\nonumber
&=\sum_{\varpi_{e_{9}}}(_{g_{0,0,0,0},\cdots,g_{e_{1},e_{2},e_{12},e_{13}},\cdots,e_{\infty,\infty,\mu_{a}+e_{1}-1,G_{Q}\mu_{b}+e_{2}-1}}^{\quad\qquad\qquad\qquad e_{9}})
\\
\label{a34} 
& \times
\Omega_{\varpi_{e_{9}}} e^{-\Lambda_{\varpi_{e_{9}}}\lambda}\lambda^{\Psi_{\varpi_{e_{9}}}},
\end{align}
where $(_{i_{1},i_{2},\cdots,i_{m}}^{\qquad i})=\frac{i!}{i_{1}!i_{2}!\cdots i_{m}!}$ symbolizes the multinomial coefficients,
$\Omega_{\varpi_{e_{9}}}=\prod_{e_{1},e_{2},e_{12},e_{13}} \lambda_{e_{13}}^{g_{e_{1},e_{2},e_{12},e_{13}}}$,
$\Psi_{\varpi_{e_{9}}}=\sum_{e_{1}}$
$\sum_{e_{2}}\sum_{e_{12}}\sum_{e_{13}} (e_{12}+e_{13})g_{e_{1},e_{2},e_{12},e_{13}}$
and
$\Lambda_{\varpi_{e_{9}}}=\sum_{e_{1}}\sum_{e_{2}}$ $\sum_{e_{12}} \sum_{e_{13}}(\mathcal{A}_{2}+\mathcal{B}_{2})g_{e_{1},e_{2},e_{12},e_{13}}$. 
For each element of $\varpi_{e_{9}}$, the sum in \eqref{a34} is to be performed, which can be defined as
\begin{align}
\nonumber
&\varpi_{e_{9}}
\\
\nonumber
&=[(g_{0,0,0,0},\cdots,g_{e_{1},e_{2},e_{12},e_{13}},\cdots,e_{\infty,\infty,\mu_{a}+e_{1}-1,G_{Q}\mu_{b}+e_{2}-1}):
\\
\nonumber
&  g_{e_{1},e_{2},e_{12},e_{13}}\in\mathbb{N},0\leq e_{1}\leq \infty,0\leq e_{2}\leq \infty, 
\\
\nonumber
& 0\leq e_{12}\leq \mu_{a}+e_{1}-1, 0\leq e_{13}\leq G_{Q}\mu_{b}+e_{2}-1;
\\
\label{a35}
& \sum_{e_{1},e_{2},e_{12},e_{13}} g_{e_{1},e_{2},e_{12},e_{13}}=e_{9}].
\end{align}
%Note that, the number of tuples in \eqref{a34} is $\frac{\left(\mu _a+e_5\right)}{2} \left(r_2+1\right) \left(r_2+2\right)( \frac{r_2}{3} + \mu _b G_Q)$ and sum of those tuple is $e_{9}$. 
Finally, substituting \eqref{a34} into \eqref{a32}, we get
\begin{align}
\nonumber
f_{\lambda_{min}}(\lambda)&=\sum_{e_{8}=0}^{Q-1}\sum_{e_{9}=0}^{P+Pe_{8}-1}\sum_{\varpi_{e_{9}}}\Big(\sum_{s_{2}=0}^{\infty}\sum_{e_{4}=0}^{\infty}\sum_{e_{10}=0}^{G_{Q}\mu_{b}+e_{4}-1}\mathcal{J}_{1}\lambda^{\mathcal{J}_{3}} 
\\
\label{a36}
& +\sum_{s_{3}=0}^{\infty}\sum_{e_{5}=0}^{\infty}\sum_{e_{11}=0}^{\mu_{a}+s_{3}-1}\mathcal{J}_{2}\lambda^{\mathcal{J}_{4}}\Big)e^{-\mathcal{J}_{5}\lambda},
\end{align}
where 
\\
$\lambda_{\varpi_{e_{9}}}=\big(_{g_{0,0,0,0},\cdots,g_{e_{1},e_{2},e_{12},e_{13}},\cdots,e_{\infty,\infty,\mu_{a}+e_{1}-1,G_{Q}\mu_{b}+e_{2}-1}}^{\quad\qquad\qquad\qquad e_{9}}\big)$
\\
$\times \Omega_{\varpi_{e_{9}}}$,  $\mathcal{J}_{1}=PQ\lambda_{e_{9}}\lambda_{e_{10}}$ $\lambda_{\varpi_{e_{9}}}$, $\mathcal{J}_{2}=PQ\lambda_{e_{9}}$ $\lambda_{e_{11}}\lambda_{\varpi_{e_{9}}}$,
$\mathcal{J}_{3}=\mu_{a}+\Psi_{\varpi_{e_{9}}}+s_{2}+e_{10}-1$, $\mathcal{J}_{4}=G_{Q}\mu_{b}+\Psi_{\varpi_{e_{9}}}+e_{5}+e_{11}-1$, and $\mathcal{J}_{5}=\mathcal{A}_{2}+\mathcal{B}_{2}+\Lambda_{\varpi_{e_{9}}}$.

%*************************************************************************************************
\subsection{Modeling of Eavesdropper Channels}
In order to perform the secrecy analysis assuming the worst possible case (i.e. maximum strength of the eavesdroppers), we herein, consider maximum SNR among $W$ eavesdroppers as denoted by $\lambda_{max}=max_{1<c<W}\lambda_{c}^{*}$. Similar to the multicast channels, $\lambda_{1}^{*}$, $\lambda_{2}^{*}$, $\ldots$, $\lambda_{W}^{*}$ are independent, and by means of order statistics, the PDF of $\lambda_{max}$ is defined as \cite{nafis2021secrecy}
%************************************************
\begin{align}
\label{a37}
f_{\lambda_{max}}(\lambda)=W f_{*,c}(\lambda)[F_{*,c}(\lambda)]^{W-1}.
\end{align}
Hence, substituting \eqref{a27} and \eqref{a28} into \eqref{a37}, and after some mathematical manipulation, $f_{\lambda_{max}}(\lambda)$ is obtained as
%***********************************************
\begin{align}
\nonumber
f_{\lambda_{max}}(\lambda)&=PW\biggl[\sum_{s_{4}=0}^{\infty}\sum_{e_{6}=0}^{\infty}\frac{\lambda_{e_{6}}\lambda^{\mu_{a}+s_{4}-1}}{e^{\mathcal{A}_{2}\lambda}}\Gamma(G_{W}\mu_{c}+e_{6},\mathcal{C}_{2}\lambda)
\\
\nonumber
& +\sum_{s_{5}=0}^{\infty}\sum_{e_{7}=0}^{\infty}\frac{\lambda_{e_{7}}\lambda^{G_{W}\mu_{c}+e_{7}-1}}{e^{\mathcal{C}_{2}\lambda}}\Gamma(\mu_{a}+s_{5},\mathcal{A}_{2}\lambda)\biggl]
\\
\nonumber
& \times \Big[1-\sum_{e_{3}=0}^{\infty}\sum_{e_{1}=0}^{\infty}\lambda_{e_{3}}\Gamma(\mu_{a}+e_{1},\mathcal{A}_{2}\lambda)
\\
\label{a38}
& \times \Gamma(G_{W}\mu_{c}+e_{3},\mathcal{C}_{2}\lambda)\Big]^{PW-1}.
\end{align}
Simplifying \eqref{a38} similar to \eqref{a32}, we get
%************************************************
\begin{align}
\nonumber
f_{\lambda_{max}}(\lambda)&=PW\sum_{e_{14}=0}^{PW-1}\lambda_{e_{14}}e^{-(\mathcal{A}_{2}+\mathcal{C}_{2})\lambda}[\gamma_{3}(\lambda)]^{e_{14}}
\biggl[\sum_{s_{4}=0}^{\infty}\sum_{e_{6}=0}^{\infty}
\\
\nonumber
& \times \sum_{e_{15}=0}^{G_{W}\mu_{c}+e_{6}-1} \lambda_{e_{15}}\lambda^{\mu_{a}+s_{4}+e_{15}-1}+\sum_{s_{5}=0}^{\infty}\sum_{e_{7}=0}^{\infty}
\\
\label{a39}
& \times \sum_{e_{16}=0}^{\mu_{a}+s_{5}-1}\lambda_{e_{16}}\lambda^{G_{W}\mu_{c}+e_{7}+e_{16}-1}\biggl],
\end{align}
where
$\lambda_{e_{14}}=(-1)^{e_{14}}(_{\,\,\,\, e_{14}}^{PR-1})$,
$\lambda_{e_{15}}= \frac{\lambda_{e_{6}}\Gamma(e_{6}+G_{W}\mu_{c})}{e_{15}!\mathcal{C}_{2}^{-e_{15}}}$ and
$\lambda_{e_{16}}= \frac{\lambda_{e_{7}}\Gamma(s_{5}+\mu_{a})}{e_{16}!\mathcal{A}_{2}^{-e_{16}}}$.
Here,
%************************************************
\begin{align}
\nonumber
\gamma_{3}(\lambda)&=\sum_{e_{1}=0}^{\infty}\sum_{e_{3}=0}^{\infty}\sum_{e_{17}=0}^{\mu_{a}+e_{1}-1}\sum_{e_{18}=0}^{G_{W}\mu_{c}+e_{3}-1} \lambda_{e_{18}}\lambda^{e_{17}+e_{18}}
\\
\label{a40}
& \times e^{-(\mathcal{A}_{2}+\mathcal{C}_{2})\lambda},
\end{align}
where $\lambda_{e_{18}}=\frac{\mathcal{A}_{1}\mathcal{C}_{1}\Gamma(e_{1}+\mu_{a})\mathcal{C}_{2}^{-G_{W}\mu_{c}-e_{3}+e_{18}}\Gamma(e_{3}+G_{W}\mu_{c})
}{e_{17}!e_{18}!\mathcal{A}_{2}^{\mu_{a}+e_{1}-e_{17}}}$.
\\
Implementing multinomial theorem, we get
%************************************************
\begin{align}
\label{a41}
[\gamma_{3}(\lambda)]^{e_{14}}=\sum_{\varpi_{e_{14}}}\lambda_{\varpi_{e_{14}}} e^{-\Lambda_{\varpi_{e_{14}}}\lambda}\lambda^{\Psi_{\varpi_{e_{14}}}},
\end{align}
where 
$\lambda_{\varpi_{e_{14}}}=\sum_{\varpi_{e_{14}}}$ 
\\
$\times \big(_{h_{0,0,0,0},\cdots,h_{e_{1},e_{3},e_{17},e_{18}},\cdots,e_{\infty,\infty,\mu_{a}+s_{5}-1,G_{W}\mu_{c}+e_{3}-1}}^{\qquad\qquad\qquad\qquad\qquad e_{14}}\big)\Omega_{\varpi_{e_{14}}}$,
\\
$\Omega_{\varpi_{e_{14}}}=\prod_{e_{1},e_{3},e_{17},e_{18}} \lambda_{e_{18}}^{h_{e_{1},e_{3},e_{17},e_{18}}}$, $\Psi_{\varpi_{e_{14}}}=\sum_{e_{1}}\sum_{e_{3}}$
\\
$\sum_{e_{17}}\sum_{e_{18}} (e_{17}+e_{18})
h_{e_{1},e_{3},e_{17},e_{18}}$,
$\Lambda_{\varpi_{e_{14}}}=\sum_{e_{1}}\sum_{e_{3}}\sum_{e_{17}}$ 
\\
$\sum_{e_{18}}(\mathcal{A}_{2}+\mathcal{C}_{2})h_{e_{1},e_{3},e_{17},e_{18}}$.
\\
Finally, substituting \eqref{a41} into \eqref{a39}, we obtain
%************************************************
\setcounter{equation}{40}
\begin{align}
\nonumber
f_{\lambda_{max}}(\lambda)&=\sum_{e_{14}=0}^{PW-1}\sum_{\varpi_{e_{14}}}\Big(\sum_{s_{4}=0}^{\infty}\sum_{e_{6}=0}^{\infty}\sum_{e_{15}=0}^{G_{W}\mu_{c}+e_{6}-1}\mathcal{J}_{6}\lambda^{\mathcal{J}_{8}}
\\
\label{a42}
& +\sum_{s_{5}=0}^{\infty}\sum_{e_{7}=0}^{\infty}\sum_{e_{16}=0}^{\mu_{a}+s_{5}-1}\mathcal{J}_{7}\lambda^{\mathcal{J}_{9}}\Big) e^{-\mathcal{J}_{10}\lambda},
\end{align}
where
%************************************************
$\mathcal{J}_{6}=PW\lambda_{e_{14}}\lambda_{e_{15}}\lambda_{\varpi_{e_{14}}}$,
$\mathcal{J}_{7}=PW\lambda_{e_{14}}\lambda_{e_{16}}\lambda_{\varpi_{e_{14}}}$,
\\
$\mathcal{J}_{8}=\mu_{a}+\Psi_{\varpi_{e_{14}}}+s_{4}+e_{15}-1$,
$\mathcal{J}_{9}=G_{W}\mu_{c}+\Psi_{\varpi_{e_{14}}}+e_{7}+e_{16}-1$ and $\mathcal{J}_{10}=\mathcal{A}_{2}+\mathcal{C}_{2}+\Lambda_{\varpi_{e_{14}}}$.

%*************************************************************************************************
\section{Performance Metrics}
\label{s2}
In the following parts, we utilize $f_{\lambda_{min}}(\lambda)$ and $f_{\lambda_{max}}(\lambda)$ of \eqref{a36} and \eqref{a42} to derive analytical expressions of three performance metrics i.e. SOPM, PNSMC, and ESMC.

%************************************************
\setcounter{eqnback}{\value{equation}}
\setcounter{equation}{43}
\begin{figure*}[!ht]
\begin{align}
\nonumber
P_{out}(\xi_{s})
&=1-\sum_{e_{8}=0}^{Q-1}\sum_{e_{9}=0}^{P+Pe_{8}-1}\sum_{\varpi_{e_{9}}}\sum_{e_{14}=0}^{PW-1}\sum_{\varpi_{e_{14}}}\Bigg[\sum_{s_{2}=0}^{\infty}\sum_{e_{4}=0}^{\infty}\sum_{e_{10}=0}^{G_{Q}\mu_{b}+e_{4}-1}
 \sum_{f_{2}=0}^{\mathcal{J}_{3}}\sum_{f_{4}=0}^{f_{2}} \bigg[\sum_{s_{4}=0}^{\infty}\sum_{e_{6}=0}^{\infty}\sum_{e_{15}=0}^{G_{W}\mu_{c}+e_{6}-1}\frac{\mathcal{J}_{6}\omega_{5}(\mathcal{J}_{8}+f_{4})!}{(\mathcal{J}_{10}+\mathcal{J}_{5}q_{\psi_{s}})^{\mathcal{J}_{8}+f_{4}+1}}
\\
\nonumber
&+\sum_{s_{5}=0}^{\infty}\sum_{e_{7}=0}^{\infty}\sum_{e_{16}=0}^{\mu_{a}+s_{5}-1} \frac{\mathcal{J}_{7}\omega_{5}(\mathcal{J}_{9}+f_{4})!}{(\mathcal{J}_{10}+\mathcal{J}_{5}q_{\psi_{s}})^{\mathcal{J}_{9}+f_{4}+1}}\bigg]-\sum_{s_{3}=0}^{\infty}\sum_{e_{5}=0}^{\infty}
\sum_{e_{11}=0}^{\mu_{a}+s_{3}-1}\sum_{f_{3}=0}^{\mathcal{J}_{4}}\sum_{f_{5}=0}^{f_{3}} \bigg[\sum_{s_{4}=0}^{\infty}\sum_{e_{6}=0}^{\infty}\sum_{e_{15}=0}^{G_{W}\mu_{c}+e_{6}-1}
\\
\label{a48}
& \times \frac{\mathcal{J}_{6}\omega_{6}(\mathcal{J}_{8}+f_{5})!}{(\mathcal{J}_{10}+\mathcal{J}_{5}q_{\psi_{s}})^{\mathcal{J}_{8}+f_{5}+1}}+\sum_{s_{5}=0}^{\infty}\sum_{e_{7}=0}^{\infty}\sum_{e_{16}=0}^{\mu_{a}+s_{5}-1}
\frac{\mathcal{J}_{7}\omega_{6}(\mathcal{J}_{9}+f_{5})!}{(\mathcal{J}_{10}+\mathcal{J}_{5}q_{\psi_{s}})^{\mathcal{J}_{9}+f_{5}+1}}\bigg]\Bigg].
\end{align}
\hrulefill
\end{figure*}
%************************************************

\subsection{SOPM Analysis}
Signifying the target secrecy rate by $\xi_{s}$, and the secrecy multicast capacity as $\mathcal{C}_{s,m}$ \cite{badrudduza2019performance}, the SOPM is denoted as 
%************************************************
\begin{align}
\setcounter{eqnback}{\value{equation}} \setcounter{equation}{41}
\nonumber
P_{out}(\xi_{s})&=\Pr(\mathcal{C}_{s,m}<\xi_{s})
\\
\label{a46}
&=1-\int_{0}^{\infty}\int_{\psi_{s}}^{\infty}f_{\lambda_{min}}(\lambda_{b}^{*})f_{\lambda_{max}}(\lambda_{c}^{*})d\lambda_{b}^{*}d\lambda_{c}^{*},
\end{align}
where $\psi_{s}=2^{\xi_{s}}(1+\lambda_{c}^{*})-1$ and $\xi_{s}>0$. This definition specifies that, reliable transmission is achievable only if $\mathcal{C}_{s,m}>\xi_{s}$, otherwise the security can not be guaranteed. Substituting \eqref{a36} and \eqref{a42} into \eqref{a46}, we get
\begin{align}
\setcounter{eqnback}{\value{equation}} \setcounter{equation}{42}
\nonumber
& P_{out}(\xi_{s})
\\
\nonumber
&=\int_{0}^{\infty}\int_{\psi_{s}}^{\infty}\sum_{e_{8}=0}^{Q-1}\sum_{e_{9}=0}^{P+Pe_{8}-1}\sum_{\varpi_{e_{9}}}\Big(\sum_{s_{2}=0}^{\infty}\sum_{e_{4}=0}^{\infty}\sum_{e_{10}=0}^{G_{Q}\mu_{b}+e_{4}-1}\mathcal{J}_{1}\lambda_{b}^{{*}^{\mathcal{J}_{3}}}
\\
\nonumber
& + \sum_{s_{3}=0}^{\infty}\sum_{e_{5}=0}^{\infty} \sum_{e_{11}=0}^{\mu_{a}+s_{3}-1}\mathcal{J}_{2}\lambda_{b}^{{*}^{\mathcal{J}_{4}}}\Big)e^{-\mathcal{J}_{5}\lambda_{b}^{*}} \sum_{e_{14}=0}^{PW-1}\sum_{\varpi_{e_{14}}}\Big(\sum_{s_{4}=0}^{\infty}\sum_{e_{6}=0}^{\infty}
\\
\nonumber
& \times \sum_{e_{15}=0}^{G_{W}\mu_{c}+e_{6}-1}\mathcal{J}_{6}\lambda_{c}^{{*}^{\mathcal{J}_{8}}}+\sum_{s_{5}=0}^{\infty}\sum_{e_{7}=0}^{\infty}\sum_{e_{16}=0}^{\mu_{a}+s_{5}-1}\mathcal{J}_{7}\lambda_{c}^{{*}^{\mathcal{J}_{9}}}\Big)
\\
\label{a47}
& \times e^{-\mathcal{J}_{10}\lambda_{c}^{*}}d\lambda_{c}^{*}d\lambda_{b}^{*}.
\end{align}
Now, performing integration making use of \cite[(eq.~3.351.2, 3.351.3)]{GR:07:Book} in \eqref{a47}, the closed form expression for the SOPM is given in \eqref{a48}, where
$p_{\psi_{s}}=2^{\xi_{s}}-1$, $q_{\psi_{s}}=2^{\xi_{s}}$,
 $\omega_{5}=\frac{ \mathcal{J}_{1}\mathcal{J}_{3}!(_{f_{4}}^{f_{2}})p_{\psi_{s}}^{f_{2}-f_{4}}}{f_{2}!\mathcal{J}_{5}^{\mathcal{J}_{3}-f_{2}+1}q_{\psi_{s}}^{-f_{4}}e^{\mathcal{J}_{5}p_{\psi_{s}}}}$ and
 $\omega_{6}=\frac{ \mathcal{J}_{2}\mathcal{J}_{4}!(_{f_{5}}^{f_{3}})p_{\psi_{s}}^{f_{3}-f_{5}}}{f_{3}!\mathcal{J}_{5}^{\mathcal{J}_{4}-f_{3}+1}q_{\psi_{s}}^{-f_{5}}e^{\mathcal{J}_{5}p_{\psi_{s}}}}$.
 
\subsubsection*{Significance of SOPM Expression}

{\color{black} It can be seen that, \eqref{a48} comprises of all the system parameters of the proposed network which helps to evaluate the secrecy outage behaviour of the proposed system. Hence \eqref{a48} can explain and quantify the secrecy trade-off in terms of the secrecy outage probability utilizing opportunistic relaying mechanism. Besides, \eqref{a48} also exhibits generic characteristics which helps the design engineers to model more practical networks.} Note that, as a special case of the proposed network with $\kappa_{a} = K$, $\mu_{a}=1$ and $m_{a}=m$, we obtain Shadowed Rician fading distribution for satellite links and for $\kappa_{b}=\kappa_{c}=0$, $\mu_{b}=\mu_{c}=m$ and $m_{b}=m_{c}\rightarrow \infty$, we obtain Nakagami-$m$ fading distribution for terrestrial links. For this particular case our results with \eqref{a48} totally matches with \cite[eq.~45]{bankey2019physical}. Likewise, for a special case with ($\kappa_{b} = \kappa_{c} = \kappa_{a} = 0$, $\mu_{b} = \mu_{c} = \mu_{a} = m$, and $m_{b} = m_{c} = m_{a} \rightarrow \infty$), our results with \eqref{a48} can be shown similar to uncorrelated (correlation coefficient $\rightarrow$ 0 ) Nakagami-$m$ fading channel in \cite{badrudduza2020enhancing}.

%*************************************************************************************************
\subsection{PNSMC Analysis}
{\color{black}The PNSMC can be expressed as
%************************************************
\setcounter{eqnback}{\value{equation}} \setcounter{equation}{44}
\begin{align}
\label{a43}
\Pr(\mathcal{C}_{s,m}>0)=\int_{0}^{\infty}\int_{0}^{\lambda_{b}^{*}}f_{\lambda_{min}}(\lambda_{b}^{*})f_{\lambda_{max}}(\lambda_{c}^{*})d\lambda_{c}^{*}d\lambda_{b}^{*}.
\end{align}
Substituting \eqref{a36} and \eqref{a42} into \eqref{a43}, $\Pr(\mathcal{C}_{s,m}>0)$ can be easily derived. But PNSMC can also be derived from the SOPM expression as the following.
\begin{align}
\label{a45}
    \Pr(\mathcal{C}_{s,m}>0)=1-P_{out}(\xi_{s})|_{\xi_{s}=0}.
\end{align}
Hence we first substitute $P_{out}(\xi_{s})$ from \eqref{a48} into \eqref{a45}, then set $\xi_{s}=0$, and finally obtain the expression of PNSMC.}

% \subsubsection*{Significance of PNSMC Expression}

% {\color{black} Note that, in \eqref{a45}, the PNSMC is expressed in terms of all system parameters that significantly affects the secrecy performance of our proposed system model. Hence, how the number of destination receivers, eavesdroppers, fading, shadowing impose detrimental impacts on system's secrecy performance and how this detrimental impact can be overcome utilizing opportunistic relaying strategy can easily be quantified using  \eqref{a45}. Besides, \eqref{a45} also explains the generic nature of our proposed model.} It is noteworthy that, for a special case with ($\kappa_{b} = \kappa_{c} = \kappa_{a} = 0$, $\mu_{b} = \mu_{c} = \mu_{a} = m$, and $m_{b} = m_{c} = m_{a} \rightarrow \infty$), our results using \eqref{a45} completely matches with the results obtained with Nakagami-$m$ fading channel (considering uncorrelated case) in \cite{badrudduza2020enhancing}. Correspondingly, for the special case with $P=0$, $Q=W=1$, $\kappa_{b}=\kappa_{c}=\kappa_{a} = K$, $\mu_{b}=\mu_{c}=\mu_{a}= 1$ and $m_{b}=m_{c}=m_{a}= m$, our analysis implementing \eqref{a45} is in a good agreement with \cite[eq.~25]{an2018secrecy}.

%*************************************************************************************************
\subsection{ESMC Analysis}
%%%%%%%%%%%%%%%%%%%%%%%%%%    49    %%%%%%%%%%%%%%%%%%%%%%%%%%%%%%%
\setcounter{eqnback}{\value{equation}} 
\setcounter{equation}{47}
\begin{figure*}[!ht]
\begin{align}
\nonumber
\langle \mathcal{C}_{s,m}\rangle
& =\sum_{e_{8}=0}^{Q-1}\sum_{e_{9}=0}^{P+Pe_{8}-1}\sum_{\varpi_{e_{9}}}
\Biggl[ \sum_{s_{2}=0}^{\infty}\sum_{e_{4}=0}^{\infty}\sum_{e_{10}=0}^{G_{Q}\mu_{b}+e_{4}-1}\biggl[\sum_{d_{2}=0}^{\mathcal{J}_{3}}\frac{D_{5}}{ln(2)}
\frac{(-1)^{D_{1}-1}Ei(-\mathcal{J}_{5})}{(\frac{1}{\mathcal{J}_{5}})^{D_{1}}e^{-\mathcal{J}_{5}}} +\sum_{d_{3}=1}^{D_{1}}\frac{(d_{3}-1)!}{(-\frac{1}{\mathcal{J}_{5}})^{D_{1}-d_{3}}}\biggl]
\\
\nonumber
& +\sum_{s_{3}=0}^{\infty}\sum_{e_{5}=0}^{\infty} \sum_{e_{11}=0}^{\mu_{a}+s_{3}-1}\biggl[\sum_{d_{4}=0}^{\mathcal{J}_{4}}\frac{D_{6}}{ln(2)}\frac{(-1)^{D_{2}-1}Ei(-\mathcal{J}_{5})}{(\frac{1}{\mathcal{J}_{5}})^{D_{2}}e^{-\mathcal{J}_{5}}}+\sum_{d_{5}=1}^{D_{2}}
\frac{(d_{5}-1)!}{(-\frac{1}{\mathcal{J}_{5}})^{D_{2}-d_{5}}}\biggl]\Biggl] 
\\
\nonumber
& -\sum_{e_{14}=0}^{PW-1}\sum_{\varpi_{e_{14}}}\Biggl[\sum_{s_{4}=0}^{\infty}\sum_{e_{6}=0}^{\infty}\sum_{e_{15}=0}^{G_{W}\mu_{c}+e_{6}-1} \biggl[\sum_{d_{6}=0}^{\mathcal{J}_{8}}\frac{D_{7}}{ln(2)}\frac{(-1)^{D_{3}-1}Ei(-\mathcal{J}_{10})}{(\frac{1}{\mathcal{J}_{10}})^{D_{3}}e^{-\mathcal{J}_{10}}} +\sum_{d_{7}=1}^{D_{3}}\frac{(d_{7}-1)!}{(-\frac{1}{\mathcal{J}_{10}})^{D_{3}-d_{7}}}\biggl]
\\
\label{a50}
&- \sum_{s_{5}=0}^{\infty} \sum_{e_{7}=0}^{\infty}\sum_{e_{16}=0}^{\mu_{a}+s_{5}-1}\biggl[\sum_{d_{8}=0}^{\mathcal{J}_{9}}\frac{D_{8}}{ln(2)}\frac{(-1)^{D_{4}-1}Ei(-\mathcal{J}_{10})}{(\frac{1}{\mathcal{J}_{10}})^{D_{4}}e^{-\mathcal{J}_{10}}} +\sum_{d_{9}=1}^{D_{4}}\frac{(d_{9}-1)!}{(-\frac{1}{\mathcal{J}_{10}})^{D_{4}-d_{9}}}\biggl]\Biggl].
\end{align}
\hrulefill
\end{figure*}
%%%%%%%%%%%%%%%%%%%%%%%%%%%%%%%%%%
%%%%%%%%%%%%%%%%%%%%%%%%%%%%%%%%%%%%
The ESMC can be defined as 
\setcounter{eqnback}{\value{equation}} \setcounter{equation}{46}
\begin{align}
\nonumber
\langle \mathcal{C}_{s,m}\rangle&=\int_{0}^{\infty}\log_2(1+\lambda_{b}^{*})f_{\lambda_{min}}(\lambda_{b}^{*})d\lambda_{b}^{*}
\\
\label{a49}
& -\int_{0}^{\infty} \log_2(1+\lambda_{c}^{*}) f_{\lambda_{max}}(\lambda_{c}^{*})d\lambda_{c}^{*}.
\end{align}
Replacing \eqref{a36} and \eqref{a42} into \eqref{a49} and integrating by making use of \cite[eq.~4.222.8]{GR:07:Book}, the novel expression for the ESMC is exhibited in \eqref{a50}, where $D_{1}=\mathcal{J}_{3}-d_{2}$, $D_{2}=\mathcal{J}_{4}-d_{4}$, $D_{3}=\mathcal{J}_{8}-d_{6}$, $D_{4}=\mathcal{J}_{9}-d_{8}$, $D_{5}=\frac{\mathcal{J}_{1}\mathcal{J}_{3}!}{D_{1}!\mathcal{J}_{5}^{\mathcal{J}_{3}+1}}$, $D_{6}=\frac{\mathcal{J}_{2}\mathcal{J}_{4}!}{D_{2}!\mathcal{J}_{5}^{\mathcal{J}_{4}+1}}$, $D_{7}=\frac{\mathcal{J}_{6}\mathcal{J}_{8}!}{D_{3}!\mathcal{J}_{10}^{\mathcal{J}_{8}+1}}$, and $D_{8}=\frac{\mathcal{J}_{7}\mathcal{J}_{9}!}{D_{4}!\mathcal{J}_{10}^{\mathcal{J}_{9}+1}}$.

\subsubsection*{Significance of ESMC Expression}

{\color{black} It is clear that how do the physical properties of the channels as well as the system parameters affects the secrecy capacity, can be easily quantified with \eqref{a50}. Besides, how does the cooperative diversity provided by the relays help to enhance secrecy capacity in spite of harsh channel conditions can also be evaluated easily. Since \eqref{a50} is a generalized expression, it also represents secrecy analysis over several classical models.} It is noted that, as a special case of Shadowed Rician fading distribution with ($P=0$, $Q=W=1$ $\kappa_{b}=\kappa_{c}=\kappa_{a} = K$, $\mu_{b}=\mu_{c}=\mu_{a}=1$ and $m_{b}=m_{c}=m_{a} \rightarrow m$), our obtained results with \eqref{a50} are equivalent with the analysis in \cite[eq.~51]{an2018secrecy}. In parallel, utilizing \eqref{a50}, another special scenario i.e. Shadowed Rician fading distribution for satellite links and Rayleigh fading distribution ($Q=W=1$, $\kappa_{b}=\kappa_{c}=\kappa_{a} = 0$, $\mu_{b}=\mu_{c}=\mu_{a}= 1$ and $m_{b}=m_{c}=m_{a}$$\rightarrow$ $\infty$) for terrestrial links can be shown as a special case of our model \cite{guo2018physical}.
%*************<END-EQUATION>************

%%%%%%%%%%%%%%%%<SECTION>%%%%%%%%%%%%%%%
\section{Numerical Results}
\label{s3}
In this section, the numerical results concerning the expressions of SOPM, PNSMC, and ESMC as given in \eqref{a48}, \eqref{a45}, and \eqref{a50}, respectively are demonstrated graphically in order to gain some useful insights regarding the enhancement of security by taking the advantages of the physical properties of the propagation medium. Since the infinite series converges rapidly after a few terms, we take the first 25 terms to obtain the numerical results. {\color{black}Additionally, the validity of the analytical expressions as described in the previous sections are justified by Monte-Carlo (MC) simulations.} \footnote {{\color{black} To accomplish this task, we generate $\kappa-\mu$ shadowed random variables in MATLAB and $10^{6}$ realizations of the channels are averaged to calculate each value of the secrecy parameters. The analytical results are shown using markers and corresponding simulation results are shown using solid lines.}} {\color{black}In each figure, we observe a close agreement between the analytical and simulation results which clearly justifies the validity of our analysis.}

%===============<FIGURE>================
\begin{figure}[!ht]
\vspace{-30mm}
    \centerline{\includegraphics[width=0.55\textwidth]{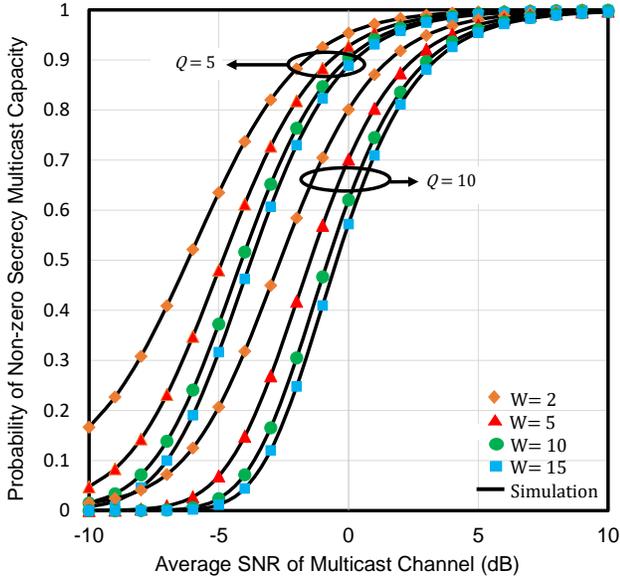}}
        \vspace{-25mm }
    \caption{{\color{black}The PNSMC versus $\bar{\lambda}_{ab}$ for selected values of $Q$ and $W$ when $\bar{\lambda}_{ac}= -10 dB$,$G_{Q}=G_{W}=2$,$\mu_{a}=\mu_{b}=\mu_{c}=2$,$\kappa_{a}=\kappa_{b}=\kappa_{c}=2$, and $m_{a}=m_{b}=m_{c}=\infty$.}}
    \label{fig:2}
\end{figure}
%=============<END-FIGURE>==============
Figure \ref{fig:2} presents the PNSMC against $\bar{\lambda}_{ab}$ in which the impact of the multicast receivers ($Q$) and eavesdroppers ($W$) are illustrated. We assume two scenarios considering $Q=$ 5 and 10. Both the scenarios clearly dictate that the PNSMC decreases with increasing values of $Q$, and $W$. In the multicast scenario, due to the increase in $Q$, the bandwidth per user is reduced which causes a reduction in the received SNR at the user node. Hence the PNSMC performance degrades remarkably which is testified in \cite{badrudduza2020enhancing}. On the contrary, increasing $W$ escalates the strength of the eavesdroppers by ensuring the maximum SNR at the eavesdropper node (since increasing $W$ indicates an increasing probability of obtaining maximum SNR at the eavesdropper channel) and thus PNSMC is degraded. It is also observed from Fig. \ref{fig:2} that the MC and analytical simulation are in a good agreement, which clearly indicates the exactness of our PNSMC expression in \eqref{a45}.

%===============<FIGURE>================
\begin{figure}[!h]
\vspace{-30mm}
     \centerline{\includegraphics[width=0.55\textwidth]{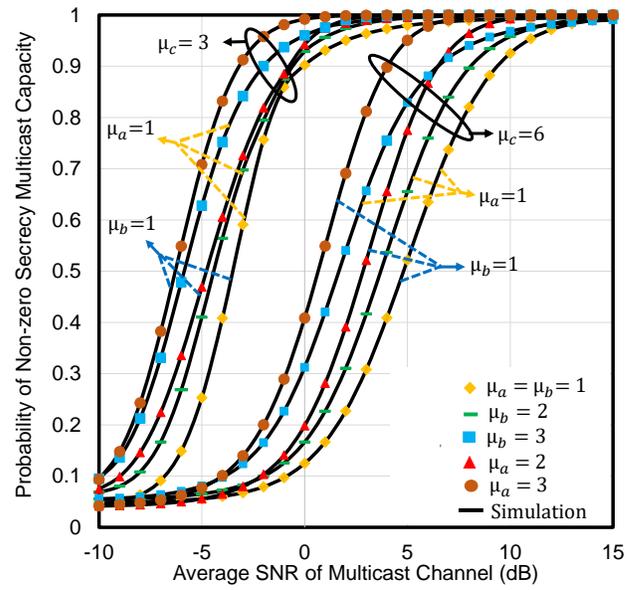}}
        \vspace{-25mm }
    \caption{{\color{black}The PNSMC versus $\bar{\lambda}_{ab}$ for selected values of $\mu_{b}$, $\mu_{a}$, and $\mu_{c}$ when $m_{a}=m_{b}=m_{c}=\infty$, $\kappa_{a}=\kappa_{b}=\kappa_{c}=1$, and $\bar{\lambda}_{ac}= 0 dB$.}}
    \label{fig:3}
\end{figure}
%=============<END-FIGURE>==============
In Fig. \ref{fig:3}, the PNSMC is plotted against $\bar{\lambda}_{ab}$ in which the effect of different number of clusters of relay channel ($\mu_{a}$), multicast channels ($\mu_{b}$) as well as eavesdropper channel ($\mu_{c}$) are shown. It is observed from the figure that the increment in the number of clusters of relay ($\mu_{a}$) and multicast channels ($\mu_{b}$) enhances the PNSMC of the proposed model. Because, with increasing values of $\mu_a$ and $\mu_b$, we are increasing the end-to-end SNR indirectly by including more incoming signals based on various diversity schemes. As a result, the total fading of the multicast channels reduces which ensures the increment of the PNSMC. Due to this similar reason, the PNSMC of the system degrades with $\mu_c$. It is noteworthy that the impact of $\mu_a$ upgrading the secrecy performance is much superior than that of $\mu_b$. The outcomes from \cite{sun2019secrecy} and \cite{sun2019physical} exhibits the same characteristics which validate our result convincingly.

%===============<FIGURE>================
\begin{figure}[!h]
\vspace{-10mm}
    \centerline{\includegraphics[width=0.75\textwidth]{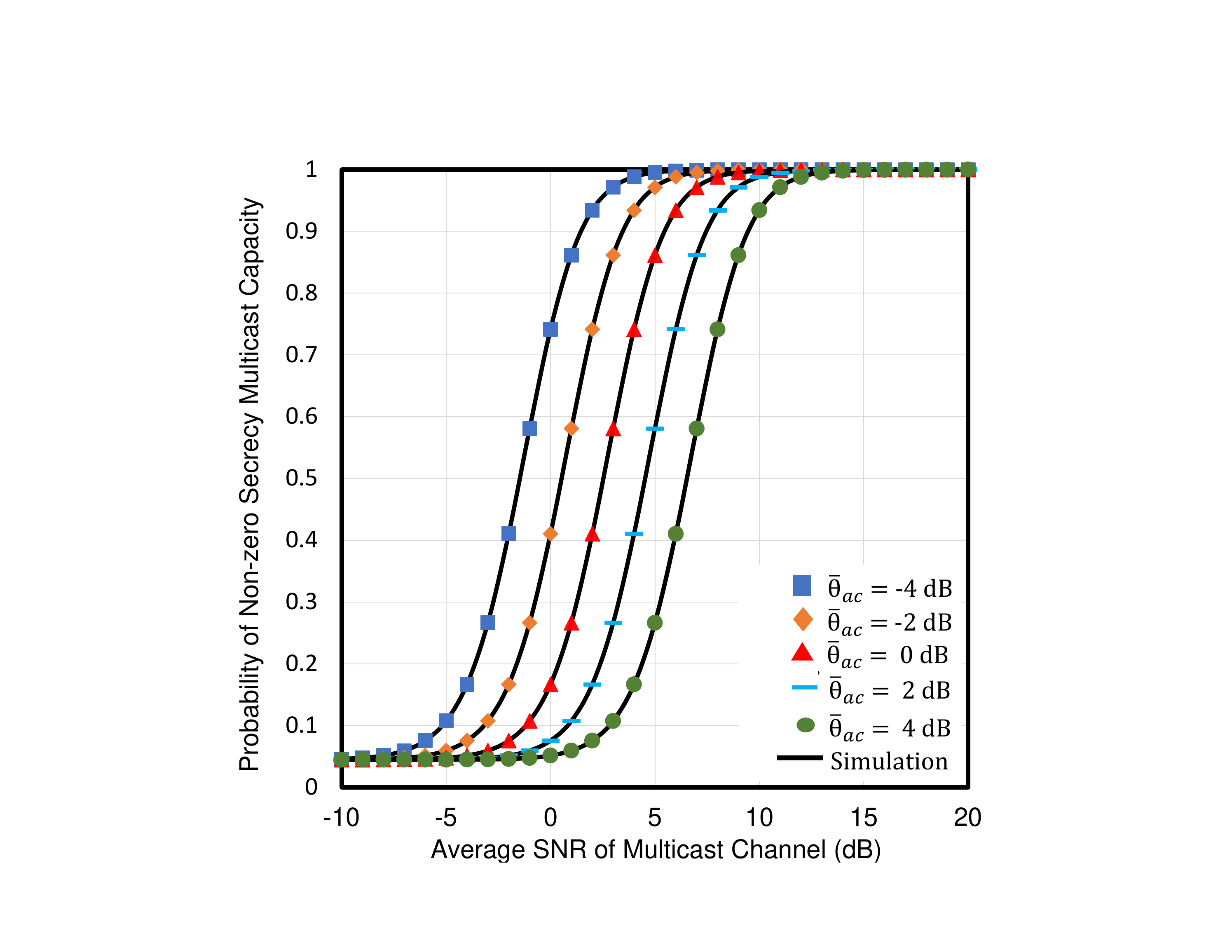}}
        \vspace{-10mm}
    \caption{{\color{black}The PNSMC versus $\bar{\lambda}_{ab}$ for selected values of $\bar{\lambda}_{ac}$ when $G_{Q}=G_{W}=2$, $\mu_{a}=\mu_{b}=\mu_{c}=2$, $\kappa_{a}=\kappa_{b}=\kappa_{c}=1$, and $m_{a}=m_{b}=m_{c}=\infty$.}}
    \label{fig:4}
\end{figure}
%=============<END-FIGURE>==============
The MC simulation and analytical results of PNSMC as a function of $\bar{\lambda}_{ab}$ is shown in Fig. \ref{fig:4}, where the outcomes due to the variation in average SNR of the eavesdropper channel ($\bar{\lambda}_{ac}$) are depicted. It is noted that the PNSMC is degraded with the increase of $\bar{\lambda}_{ac}$ because an increase in $\bar{\lambda}_{ac}$ indicates a higher SNR at the eavesdropper terminals which enhances the strength of the wiretap channels in terms of secrecy performance as shown in \cite{bhargav2016secrecy}.

%===============<FIGURE>================
\begin{figure}[!h]
\vspace{-30mm}
    \centerline{\includegraphics[width=0.55\textwidth]{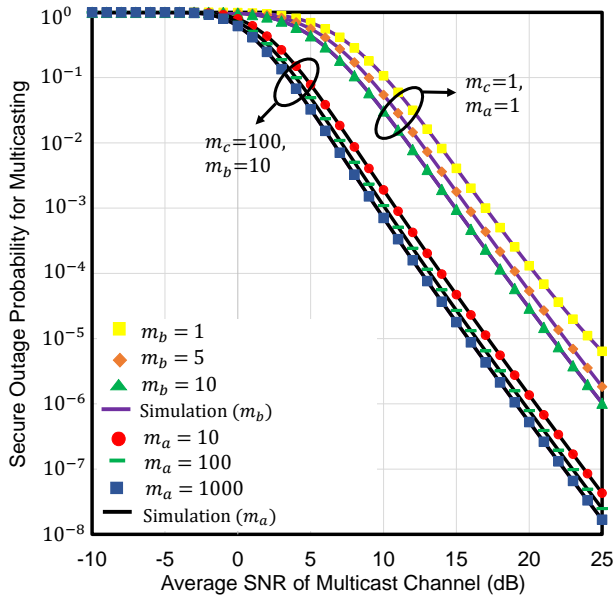}}
        \vspace{-24mm }
    \caption{{\color{black}The SOPM versus $\bar{\lambda}_{ab}$ for selected values of $m_{b}$, $m_{a}$, and $m_{c}$ when $\bar{\lambda}_{ac}= -5 dB$, $G_{Q}=G_{W}=2$, $\mu_{a}=\mu_{b}=\mu_{c}=1$, and $\kappa_{a}=\kappa_{b}=\kappa_{c}=1$.}}
    \label{fig:5}
\end{figure}
%=============<END-FIGURE>==============
Figure \ref{fig:5} demonstrates the SOPM as a function of $\bar{\lambda}_{ab}$, where the consequences of variation in the shadowing parameter for multicast channels ($m_{b}$) and relay channel ($m_{a}$) are represented. From the figure, it is recognized that increment in $m_{b}$ and $m_{a}$ minimize the shadowing effect of the multicast channel, and thus the SOPM of the model decrease. So superior secrecy performance can be achieved if the shadowing present in the multicast network is comparatively lower. Our numerical result is also confirmed by MC simulation. A similar conclusion (related to shadowing) was also drawn in \cite{sun2019secrecy}.

%===============<FIGURE>================
\begin{figure}[!h]
\vspace{-25mm}
    \centerline{\includegraphics[width=0.55\textwidth]{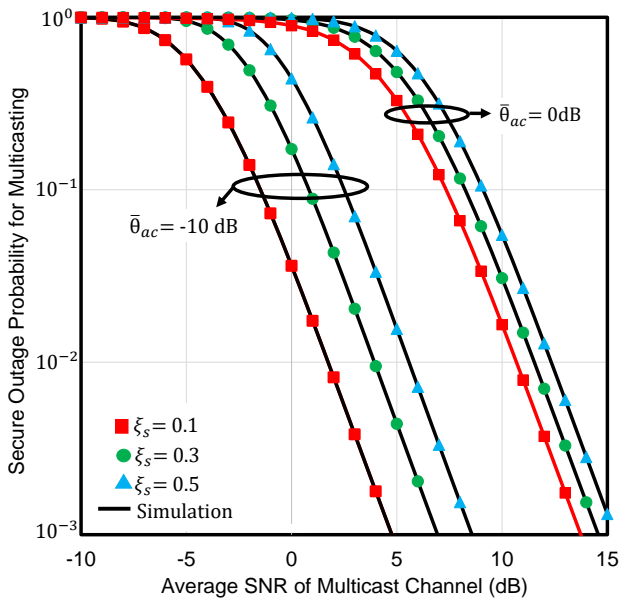}}
        \vspace{-24mm }
    \caption{{\color{black}The SOPM versus $\bar{\lambda}_{ab}$ for selected values of $\xi_{s}$ and $\bar{\lambda}_{ac}$.}}
    \label{fig:6}
\end{figure}
%=============<END-FIGURE>==============
Figure \ref{fig:6} depicts SOPM versus $\bar{\lambda}_{ab}$ to represent the effect of target secrecy rate ($\xi_{s}$) on the secure outage performance. We consider two cases herein with $\bar{\lambda}_{ac}=$ 0 and -10 dB. It is noted that the SOPM increases with $\xi_{s}$ for both cases of $\bar{\lambda}_{ac}$ as shown in \cite{badrudduza2021secrecy}. It is also noticeable that the impact of $\xi_{s}$ in case of $\bar{\lambda}_{ac}$= -10 dB is more significant than that of $\bar{\lambda}_{ac}$= 0 dB. Moreover, the SOPM performance improves when the eavesdropper channel becomes worse (i.e. $\bar{\lambda}_{ac}=$ -10 dB). A good agreement between MC and analytical outcomes proves that the SOPM derived in \eqref{a48} is accurate.

%===============<FIGURE>================
\begin{figure}[!h]
\vspace{-13mm}
     \centerline{\includegraphics[width=0.75\textwidth]{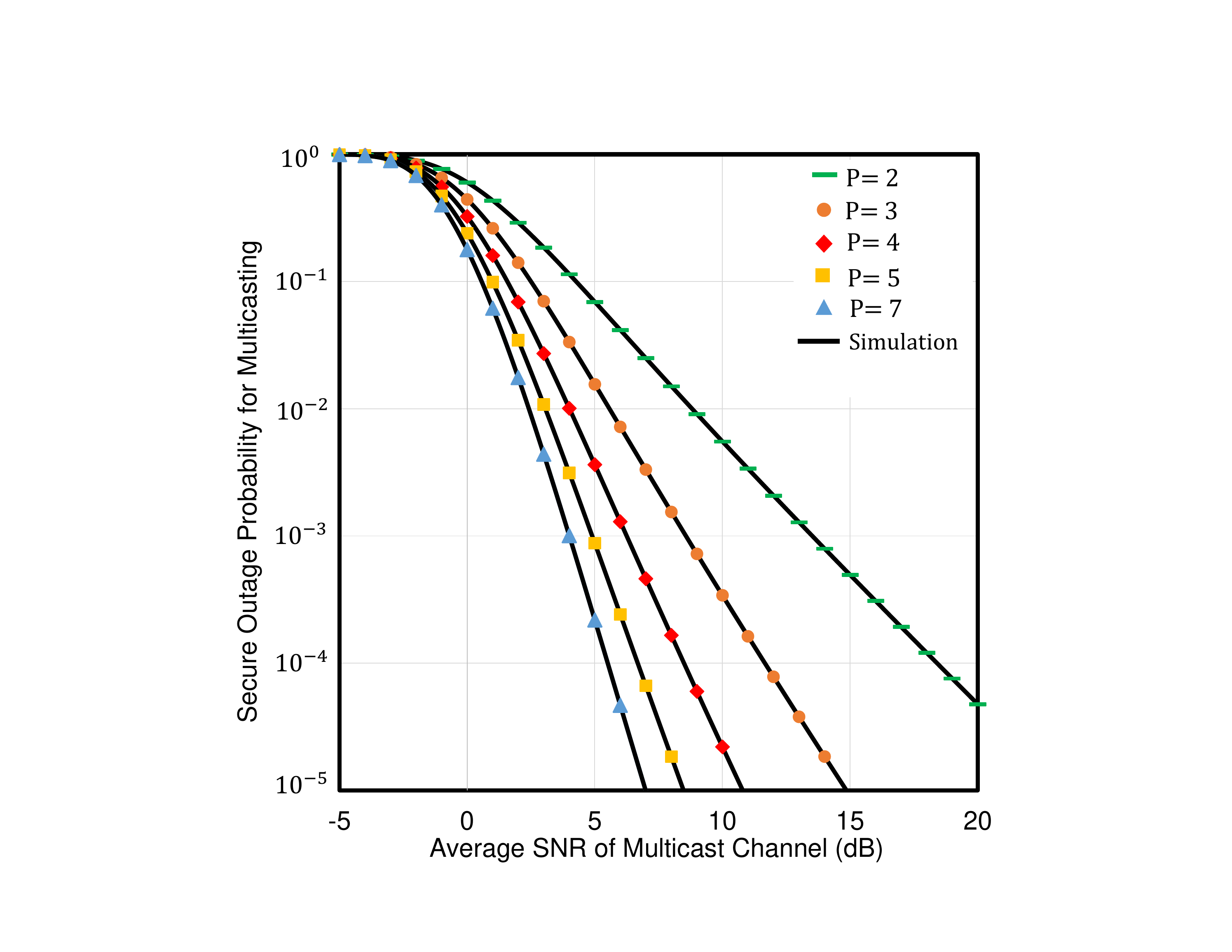}}
        \vspace{-10mm }
    \caption{{\color{black}The SOPM versus $\bar{\lambda}_{ab}$ for selected values of $P$ when $\bar{\lambda}_{ac}= -10 dB$, $G_{Q}=G_{W}=2$, $\mu_{a}=\mu_{b}=\mu_{c}=1$, $\kappa_{a}=\kappa_{b}=\kappa_{c}=1$, and  $m_{a}=m_{b}=m_{c}=\infty$.}}
    \label{fig:7}
\end{figure}
%=============<END-FIGURE>==============
In Fig. \ref{fig:7}, the SOPM is varied against $\bar{\lambda}_{ab}$ to illustrate the consequences of differing the number of relays ($P$). From the figure, it can be seen that an increment in $P$ offers a remarkable improvements in system's SOPM performance. As the number of relays are increased, the $P$ relays compete among themselves to be the best one. Additionally, the cooperative diversity provided by multiple relays also play a notable role in reducing the impacts of fading and shadowing of multicast links.

%===============<FIGURE>================
\begin{figure}[!h]
\vspace{-27mm}
   \centerline{\includegraphics[width=0.55\textwidth]{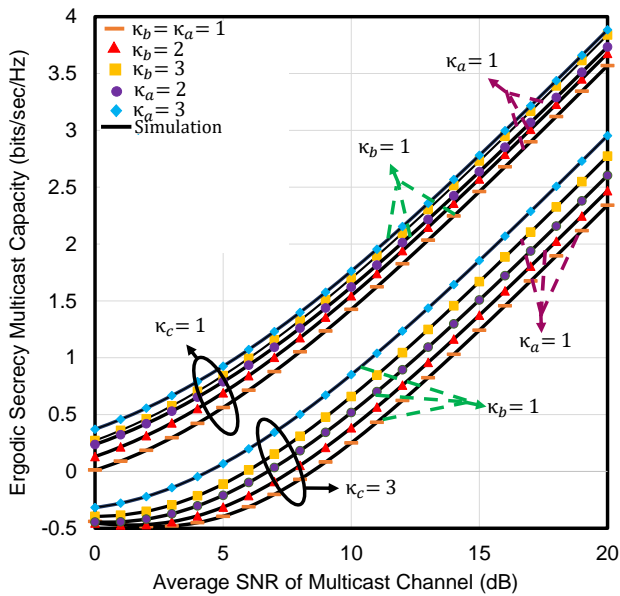}}
        \vspace{-25mm }
    \caption{{\color{black}The ESMC versus $\bar{\lambda}_{ab}$ for selected values of $\kappa_{b}$, $\kappa_{a}$, and $\kappa_{c}$ when $\mu_{a}=\mu_{b}=\mu_{c}=1$, $m_{a}=m_{b}=m_{c}=\infty$ and $\bar{\lambda}_{ac}= -5 dB$.}}
    \label{fig:8}
\end{figure}
%=============<END-FIGURE>==============
Figure. \ref{fig:8} represents ESMC with respect to $\bar{\lambda}_{ab}$ for different values of $\kappa_{a}$, $\kappa_{b}$ and $\kappa_{c}$. It is evident from the figure that both $\kappa_{a}$ and $\kappa_{b}$ increases the secrecy capacity of the proposed system. The increment of $\kappa_{a}$ and $\kappa_{b}$ enhances the dominant component as well as reduces the scattering component of the multicast channel which results in higher SNR at the receiver and ESMC of the proposed model. But, the capacity degrades sharply with $\kappa_{c}$ as it upgrades the SNR of the eavesdropper channel. Same results were obtained in \cite{sun2019secrecy}, and \cite{sun2019physical} which undoubtedly confirms our analysis. The MC simulation also has a tight agreement with the analytical result.

%===============<FIGURE>================
\begin{figure}[!h]
\vspace{-27mm}
    \centerline{\includegraphics[width=0.55\textwidth]{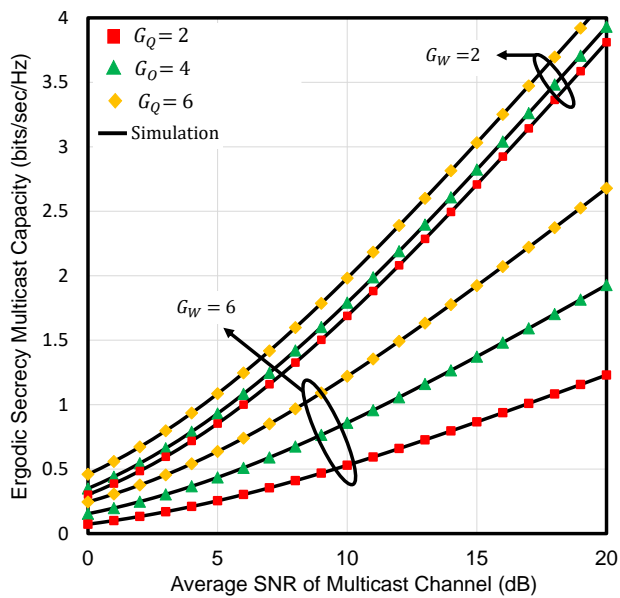}}
        \vspace{-25mm }
    \caption{{\color{black}The ESMC versus $\bar{\lambda}_{ab}$ for selected values of $G_{Q}$ and $G_{W}$ when $\bar{\lambda}_{ac}= -10 dB$, $\kappa_{a}=\kappa_{b}=\kappa_{c}=1$, $\mu_{a}=\mu_{b}=\mu_{c}=1$, and $m_{a}=m_{b}=m_{c}=\infty$.}}
    \label{fig:9}
\end{figure}
%=============<END-FIGURE>==============
In Fig. \ref{fig:9}, ESMC is plotted against $\bar{\lambda}_{ab}$ with a view to observing the impacts due to the variation in number of antennas of each user ($G_{Q}$) and eavesdropper ($G_{W}$). From the figure, it is observed that the ESMC escalates if $G_{Q}$ increases. This is because an increase in $G_{Q}$ significantly reduce the fading of multicast channels by enhancing the antenna diversity at the receiver. On the other hand, it is also noted that, ESMC degrades if $G_{W}$ is increased. In that case, the eavesdroppers are capable of overhearing more confidential messages from the multicast channels due to increase in antenna diversity at the eavesdropper terminals. Even ESMC degrades dramatically when the eavesdroppers are equipped with large number of antennas (i.e. $G_{W}$=6). Similar outcomes are also presented in \cite{srinivasan2018secrecy} which manifest the exactness of our results.

%===============<FIGURE>================
\begin{figure}[!h]
\vspace{-30mm}
   \centerline{\includegraphics[width=0.55\textwidth]{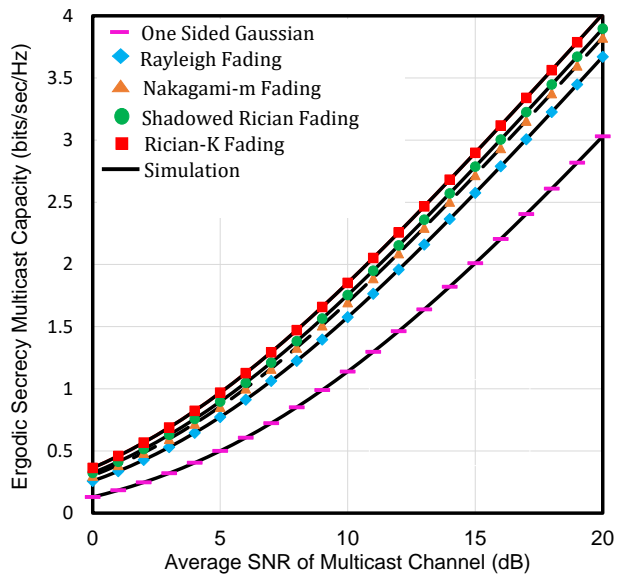}}
        \vspace{-23mm}
    \caption{{\color{black}The ESMC versus $\bar{\lambda}_{ab}$ for comparing performance of different classical fading channels as a special cases of $\kappa-\mu$ shadowed fading channel when $\bar{\lambda}_{ac}= -10 dB$, $G_{Q}=G_{W}=2$, and $P = Q = W = 2$.}}
    \label{fig:10}
\end{figure}
%=============<END-FIGURE>==============
Figure \ref{fig:10} exhibits a graphical representation in which the generic characteristics of the proposed scenario is illustrated by plotting ESMC against $\bar{\lambda}_{ab}$. Note that the secrecy analysis over generalized shadowed model using opportunistic relaying is absent in the previous studies. Moreover, simply reorienting the system parameters, we can generate some existing models as special cases which is a clear indication of superiority of our proposed work. Hence, the conclusive remarks based on the aforementioned discussion is that this novel proposed work is more purposeful than all the conventional multipath/ shadowed secure models. 

\quad

\subsection*{Comparative Analysis with Existing Related Literature}

We consider generalized distribution (i.e. $\kappa-\mu$ shadowed fading) at $S\rightarrow P$, $P\rightarrow Q$, and $P\rightarrow W$ links in the proposed dual-hop scheme that encompasses a number of well-known fading distributions which can be acquired as special cases of our model as shown in Table \ref{table:1}. Hence, it is noteworthy that the derived expressions in \eqref{a45}, \eqref{a48}, and \eqref{a50} regarding our proposed scenario are also generalized, and can be utilized to unify the secrecy performances of the mentioned channels in Table \ref{table:1}.

%%%%%%%%%%%%%%%%<SECTION>%%%%%%%%%%%%%%%
\section{Conclusions}
\label{s4}
This paper considers PLS in the dual-hop secure wireless multicast relay networks over $\kappa-\mu$ shadow-fading channels with multiple eavesdroppers. Under this scenario, secrecy enhancement is ensured by choosing the best relay among multiple relays. The effect of all the system parameters on the secrecy performance of the proposed model is thoroughly observed by deriving the exact and analytical expressions of the performance metrics, i.e., PNSMC, SOPM, and ESMC. Also, these analytical results are numerically verified with Monte-Carlo simulations. Form such analyses, it is shown that the secrecy performance of this dual-hop relay communication model mostly affected by the channel environment of the first hop, i.e., source-to-relay link, than that of the second hops, i.e., relay-to-destination and relay-to-eavesdropper links. Shadowing is an important aspect of this study. The security of this proposed scenario can be improved by increasing the shadowing effect in relay-to-eavesdropper link. The proposed generalized shadowing model with opportunistic relay operation can also be employed to improve the security of different classical fading scenarios irrespective of the harsh channel conditions in the presence of a large number of multicast receivers and eavesdroppers. The proposed model can also be extended to non-terrestrial networks where shadowing is one of the major impairments.

%%%%%%%%%%%%%%%%<SECTION>%%%%%%%%%%%%%%%
\bibliographystyle{IEEEtran}
\bibliography{IEEEabrv,main.bib}

\end{document}